\documentclass[12pt,a4paper]{article}
\pdfoutput=1

\usepackage{geometry}
\usepackage{amsmath}
\usepackage{amsmath}
\usepackage{amssymb}
\usepackage[dvips]{graphicx}
\usepackage{cite}
\usepackage{pstricks}
\usepackage{bm}
\usepackage{pbox}
\usepackage{placeins}
\usepackage{graphicx}
\usepackage{caption}
\usepackage{subcaption}
\usepackage[T1]{fontenc}
\usepackage{footnote}
\usepackage{pdfpages}
\usepackage{hhline}
\usepackage{multirow}
\usepackage{multicol}
\usepackage{enumitem}
\usepackage{multirow}
\usepackage{amsmath}
\usepackage{relsize}
 \usepackage{graphicx}
\usepackage[toc,page]{appendix}
\usepackage[english]{babel}
\allowdisplaybreaks

\usepackage{slashed}
\usepackage{cases}

\usepackage{empheq}

\renewcommand{\thetable}{\Roman{table}}

\definecolor{MyDarkBlue}{rgb}{0.1, 0.3, 0.8} 
\definecolor{SBlue}{rgb}{0.2, 0.4, 0.4} 
\definecolor{MyLightBlue}{rgb}{0.22,0.51,0.99}
\definecolor{MyGreen}{rgb}{0.0, 0.5, 0.3}
\definecolor{BrickRed}{rgb}{0.8, 0.25, 0.33}
\usepackage[colorlinks=true,linkcolor=blue,citecolor=MyDarkBlue,
urlcolor=BrickRed,bookmarksnumbered=true,bookmarksopen]{hyperref}
\hypersetup{colorlinks, citecolor=SBlue,linkcolor=MyGreen, urlcolor=MyDarkBlue}

\begin{document}
\vspace*{-0.2in}
\begin{flushright}
{ OSU-HEP-20-11}
\end{flushright}
\vspace{0.1cm}
\renewcommand{\thefootnote}{\fnsymbol{footnote}}
\begin{center}
{\Large \bf
Dark matter assisted lepton anomalous  magnetic  \\\vspace{0.1in} moments   and neutrino masses
}
\end{center}
\renewcommand{\thefootnote}{\fnsymbol{footnote}}
\begin{center}
{
{}~\textbf{Sudip Jana$^1$}\footnote{ E-mail: \textcolor{MyDarkBlue}{sudip.jana@mpi-hd.mpg.de}},
{}~\textbf{Vishnu P.K.$^2$}\footnote{ E-mail: \textcolor{MyDarkBlue}{vipadma@okstate.edu}},
{}~\textbf{Werner Rodejohann$^1$}\footnote{ E-mail: \textcolor{MyDarkBlue}{werner.rodejohann@mpi-hd.mpg.de}},
{}~\textbf{and Shaikh Saad$^2$}\footnote{ E-mail: \textcolor{MyDarkBlue}{shaikh.saad@okstate.edu}}
}
\vspace{0.5cm}
{
\\\em $^1$Max-Planck-Institut f{\"u}r Kernphysik, Saupfercheckweg 1, 69117 Heidelberg, Germany
\\
$^2$Department of Physics, Oklahoma State University, Stillwater, OK 74078, USA
} 
\end{center}

\renewcommand{\thefootnote}{\arabic{footnote}}
\setcounter{footnote}{0}
\thispagestyle{empty}

\begin{abstract}
\noindent 
We  propose a framework that addresses the origin of neutrino mass, explains the observed discrepancies in the electron and the muon anomalous magnetic moments (AMMs) data, and incorporates the dark matter (DM) relic abundance.  Both the neutrino mass and the lepton AMMs are generated at one-loop level mediated by a common set of beyond the Standard Model  (SM) states.  In this class of models, the SM is extended with  vector-like charged fermion and scalar multiplets, all odd under an imposed $\mathcal{Z}_2$ symmetry, which stabilizes the fermionic or scalar DM candidate residing in one of them. 
Two scalar multiplets appear in the AMM loops, thus allowing for different signs of their contributions, in agreement with the observed discrepancies which are of opposite sign for electron and muon. The vector-like fermions 
give rise to large new physics contributions to the lepton AMMs via chirally enhanced terms that are proportional  to  their mass. To demonstrate the viability of this framework, we perform a detailed study of a particular model for which a fit to the neutrino masses and mixing together with lepton AMMs are provided. Furthermore, DM  phenomenology and  collider signatures are explored.  
\end{abstract}

\newpage
\setcounter{footnote}{0}

{
  \hypersetup{linkcolor=black}
  \tableofcontents
}
\section{Introduction}
The origin of the neutrino mass is among the most crucial problems of the Standard Model (SM) of particle physics. On the other hand, the almost a century old dark matter (DM) problem is another tremendous  puzzle yet to be solved.  The most straightforward approach to this issue is the particle nature of the DM (for a review see Ref.\  \cite{Bertone:2004pz}). One of the many popular mechanisms for neutrino mass is the radiative one (for a recent review see Ref.\ \cite{Cai:2017jrq,Babu:2019mfe}) due to the natural accessibility of the involved particles at colliders and low energy experiments. 

There have been lots of attempts in the literature to combine these two seemingly uncorrelated issues, one of the most prominent example being the scotogenic model \cite{Ma:2006km}. In such models, the particles mediating the loop(s) that generate neutrino mass are  dark matter.  
Typically, new symmetries beyond the SM are required  to stabilize the DM and in some cases to forbid the tree-level neutrino mass contributions, for systematic studies along this line, see for example Refs.\  \cite{Restrepo:2013aga,Jana:2019mgj,Farzan:2012ev,  Simoes:2017kqb, CentellesChulia:2019xky, Jana:2019mez}. The details of these models largely depend on the nature of the imposed symmetries and the needed particle content. However, common features of these models are: (i) neutrino mass is generated via quantum corrections at a given loop order, (ii) DM candidates  naturally arise due to symmetry reasons, (iii)  owing to the loop suppression, the new physics (NP) scale can be around the TeV scale without making the Yukawa couplings unnecessarily small, which provides a way to test these models at low energies.

Aside from neutrino mass and DM, there has been a longstanding tension between the SM prediction \cite{Aoyama:2020ynm} and the experimental measured value \cite{Bennett:2006fi} of the muon anomalous magnetic moment (AMM). Additionally, the recently measured  fine-structure constant $\alpha$ using Caesium atoms with unprecedented precision \cite{Parker:2018vye} implies a deviation of the electron AMM from the SM value \cite{Aoyama:2017uqe} of opposite sign compared to the muon AMM. The experimental measurements point towards about $2.5\sigma$ and $3.7\sigma$ tensions for the electron and the muon AMMs, respectively. More precisely,  the corresponding  discrepancies  are given as 
\begin{align}
&\Delta a_{e}= a_{e}^\mathrm{exp} - a_{e}^\mathrm{SM} =-(8.7\pm 3.6)\times 10^{-13}, \label{ae}\\
&\Delta a_{\mu}= a_\mu^\mathrm{exp} - a_\mu^\mathrm{SM} = (2.79\pm 0.76)\times 10^{-9}.\label{amu}
\end{align}
Since the AMMs of light charged leptons ($a_\ell=(g-2)_\ell/2$, $\ell= e, \mu$) are measured with excellent accuracy in the experiments, and their corresponding theory values are computed with outstanding precision, these observed tensions strongly point towards physics beyond the SM. Therefore, these results recently have entertained a lot of interest in the particle physics community,  for attempts to simultaneously explain these discrepancies see Refs.\ \cite{Giudice:2012ms, Davoudiasl:2018fbb,Crivellin:2018qmi,Liu:2018xkx,Dutta:2018fge, Han:2018znu, Crivellin:2019mvj,Endo:2019bcj, Abdullah:2019ofw, Bauer:2019gfk,Badziak:2019gaf,Hiller:2019mou,CarcamoHernandez:2019ydc,Cornella:2019uxs,Endo:2020mev,CarcamoHernandez:2020pxw,Haba:2020gkr, Bigaran:2020jil, Jana:2020pxx,Calibbi:2020emz,Chen:2020jvl,Yang:2020bmh,Hati:2020fzp,Dutta:2020scq,Botella:2020xzf,Chen:2020tfr, Dorsner:2020aaz, Arbelaez:2020rbq}.  For previous analyses of non-supersymmetric models that accommodate only DM and $(g-2)_\mu$ see Refs.\ \cite{Chen:2009ata, Agrawal:2014ufa, Baek:2015fea, Belanger:2015nma, Kowalska:2017iqv, Calibbi:2018rzv, Chen:2019nud,  Calibbi:2019bay}, and for studies that make  a connection between radiative neutrino mass generation and $(g-2)_\mu$, see Refs.\  \cite{Ma:2001mr, Dicus:2001ph, Babu:2010vp, Nomura:2016jnl, Nomura:2016ask, Lee:2017ekw, Chiang:2017tai, Saad:2020ihm}.

To address both $(g-2)_e$ and $(g-2)_\mu$, NP may appear at  low scale, see for example Ref.\ \cite{Jana:2020pxx}. 
Models of these types are highly constrained from beam dump experiments, Belle and BaBar, which may eventually rule out such scenarios in the near future. We, on the other hand,  are interested in scenarios where NP emerges at heavy scale.\footnote{Both heavy and light new physics is not expected to influence the MUonE experiment \cite{Calame:2015fva}, which will directly measure the crucial hadronic vacuum polarization contribution to the muon AMM, see Refs.\ \cite{Dev:2020drf,Masiero:2020vxk}. } To incorporate large deviations for $\Delta a_\ell$ given in Eqs.\  \eqref{ae} and \eqref{amu} from heavy NP, a chirality flip of a heavy state must take place inside the loop. This can be achieved with TeV scale scalar leptoquarks \cite{Bigaran:2020jil, Dorsner:2020aaz} or vector-like fermions \cite{Crivellin:2018qmi}. These studies however, made no connection with either neutrino mass or DM issues.

In this work, we bring the issues of the origin of neutrino mass, the DM problem, and the electron and muon AMM puzzles under the same umbrella, and propose a framework for their explanations in a minimalistic approach.   In our proposed setup, the particle content of the SM is extended by three generations of vector-like fermions and three scalar multiplets. Furthermore, the model is supplemented with a $\mathcal{Z}_2$ symmetry, under which only the BSM particles are assumed to be odd. Via the propagation of these BSM multiplets, neutrino mass generation as well as new physics contributions  to the lepton anomalous magnetic moments of the correct order appear at one-loop level. Two scalar multiplets, and thus two sets of Yukawa couplings, appear in the AMM loops, thus allowing for different signs of their contributions, in agreement with the observed discrepancies which are of opposite sign for electron and muon. 
The lightest of the neutral BSM particles is stabilized by the imposed $\mathcal{Z}_2$ symmetry, which  serves as the DM candidate.  \\

The paper is build up as follows: In Section \ref{sec:frame} we address which model classes may solve the AMM discrepancies and at the same time generate neutrino mass radiatively with the same set of new multiplets. From the list of models, in Section \ref{sec:I} we perform a detailed analysis of one of them, analyzing the scalar sector, performing a fit to the AMM and neutrino mass observables, discussing the dark matter phenomenology, and outlining collider phenomenology. 
We conclude in Section \ref{sec:C}. 

\begin{figure}[t!]
\centering
\includegraphics[width=0.49\textwidth]{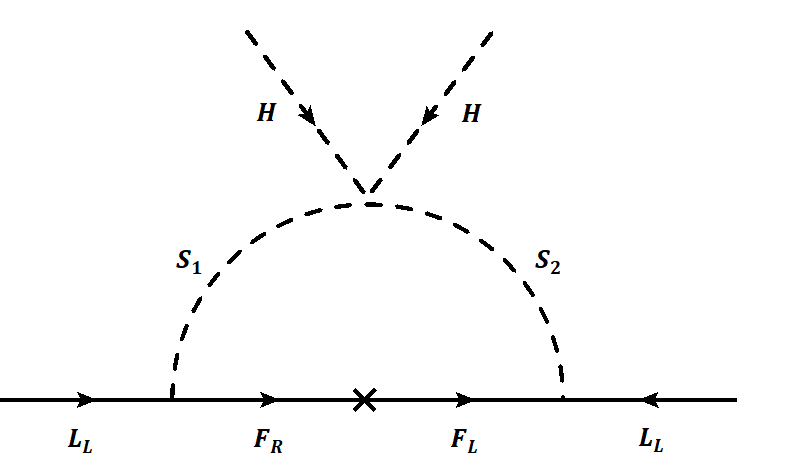}
\caption{Representative Feynman diagram for generating neutrino mass. } \label{FWA}
\end{figure}

\section{Framework\label{sec:frame}}
Due to its simplicity, we start our discussion with the scotogenic model \cite{Ma:2006km}, which employs three generations of singlet Majorana fermions\footnote{Our convention is 
$Q = I_3 + Y$.} $N (1,1,0)$ and an inert Higgs doublet $\phi (1,2,1/2)$, both odd under an imposed $\mathcal{Z}_2$ symmetry. The neutrino mass is generated at the one-loop level via the generic diagram shown in Fig.\ \ref{FWA}, with $S_1=S_2\equiv \phi$ and $F_L=F_R\equiv N$ (Majorana fermion). 
  However, a combined explanation of lepton AMMs along with reproducing realistic neutrino masses and mixings cannot be accommodated, since  the proportionality relation $|a_\mu|\propto m^2_\mu$ requires large Yukawa couplings, which would generate too large rates for charged lepton flavor violating (cLVF) processes like $\mu \to e \gamma$. 
Quantifying this tension very roughly is possible as follows. In the scotogenic model neutrino mass is given by (setting loop functions to one and assuming that all new particle masses are of order TeV)
\begin{equation}
{\cal M}_\nu \sim \frac{M_N \, y^2}{32 \pi^2} \simeq 0.3 \left(\frac{M_N}{\rm TeV} \right) 
\left(\frac{y}{10^{-5}} \right)^2\;\textrm{eV},   \label{quntifyNU}
\end{equation}
where flavor indices are ignored and $y$ is the Yukawa coupling of the singlet fermion $N$ with lepton doublets and the inert scalar doublet. This shows that to get the correct order of neutrino mass one requires $y =  \mathcal{O}(10^{-5})$.  On the other hand, the contribution to the magnetic moment is 
\begin{equation}
-\Delta a_\mu \sim m_\mu^2 \, \frac{|y|^2}{32 \pi^2 \, M_S^2} \sim 10^{-21} \left(\frac{M_S}{\rm TeV} \right)^{-2} \left(\frac{y}{10^{-5}} \right)^2 . \label{quntifyAMM}
\end{equation}
Apart from the fact that the sign of $\Delta a_\mu$ is actually not correct in the scotogenic model, these estimates show that a simultaneous explanation of neutrino mass and the anomalous magnetic moment is not possible. Moreover, the branching ratio for $\mu \to e \gamma$ provides additional constraints, namely 
\begin{equation}
{\rm BR}(\mu \to e \gamma) \sim 
\frac{3 \alpha \, y^4}{32 \pi \, G_F^2 \, M_S^4} \sim 10^{-26} \left(\frac{M_S}{\rm TeV} \right)^{-4} 
\left(\frac{y}{10^{-5}} \right)^4,  
\end{equation}
with $S$ a scalar particle of the model. Too large rates would appear 
for order one Yukawas. 
A detailed parameter scan confirms such statements \cite{Toma:2013zsa,Lindner:2016bgg}.

\begin{figure*}[t!]
\includegraphics[width=0.48\textwidth]{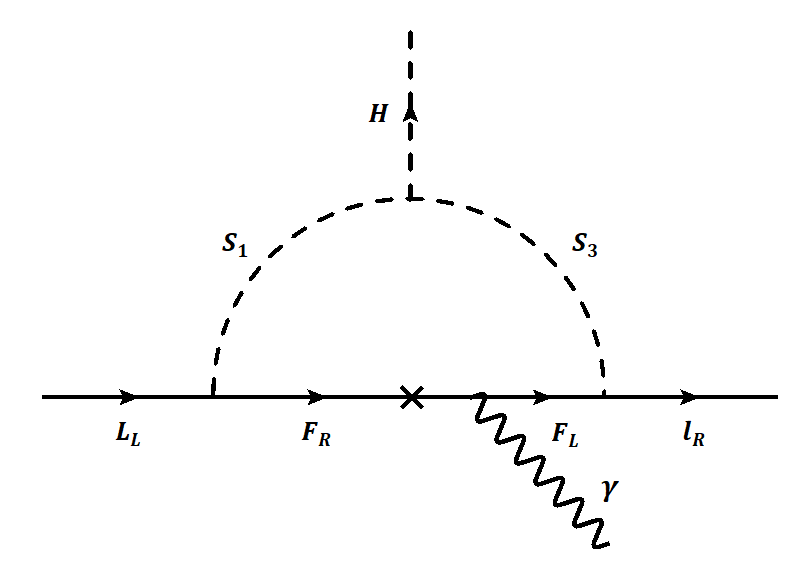}
\includegraphics[width=0.48\textwidth]{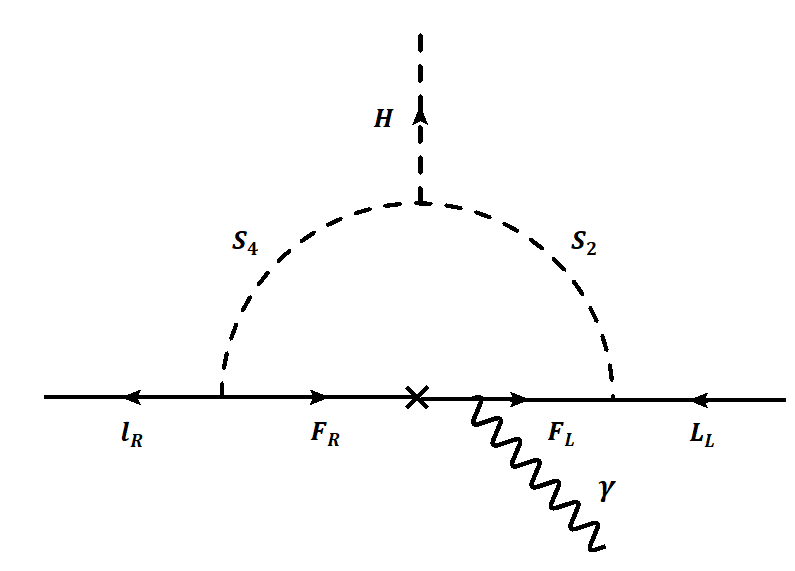}
\caption{New physics contributions to  $(g-2)_\ell$. The outgoing photon can be emitted from the internal fermion or boson line (or both) depending on the model. 
} \label{FWB}
\end{figure*}

The sign of the muon AMM could be changed by a minimal addition of one more scalar, which provides a freedom to choose the sign of the product of the Yukawa coupling in the AMM contributions. This would utilize  either of the two one-loop diagrams presented in Fig.\ \ref{FWB}.  For the scotogenic model both these diagrams are identical and the loop can be completed by introducing a singly charged scalar, that is $S_3=S_4\equiv \eta (1,1,1)$. However, the smallness of the implied AMM contribution remains,  which can be quantified as follows. The presence of $\eta^+$, with different hypercharge than the inert doublet and Yukawa coupling $y'$, allows both left-handed and right-handed charged leptons in the external legs (unlike the scotogenic model that involves only left-handed charged fermions) and provides enhanced contribution to lepton AMM that is proportional to  the mass of $N_R$. Then the formula given in Eq.\ \eqref{quntifyAMM} has the following modified form
\begin{equation}
-\Delta a_\mu \sim \frac{m_\mu }{8 \pi^2 \, M_S^2} y y' \, \theta \, M_N 
\sim 10^{-12} \left(\frac{M_{\rm NP}}{\rm TeV} \right)^{-1} 
\left(\frac{y}{10^{-5}} \right) y'\, \theta \,,  
\end{equation}
where $\theta \le 1$ represents the mixing angle between the two singly charged states, $M_{\rm NP}$ is common new physics scale of the new particles, 
and in the second line we have assume the dominance of one of the terms towards lepton AMM to maximize the effect. 
This implies that even with $y'_\ell \sim 1$, much higher values than  $y\sim 10^{-5}$ are required to explain the AMM, which would be in  conflict with neutrino mass and cLFV for TeV scale new particles.  Such correlations can be avoided if the Yukawa coupling $y$ does not participate in explaining $(g-2)_\ell$. This is precisely what we try to achieve in an economical fashion  within our framework. 

The same conclusion can be reached for any similar model with Majorana fermions running in the loop in Fig.\ \ref{FWA} that transform non-trivially under the $SU(2)_L$ group,  for example $F\sim (1,3,0)$ (for this choice, one again gets $S_1=S_2\equiv \phi$).         \\

The above arguments are also changed if hypercharged vector-like Dirac fermions instead of Majorana fermions are introduced.  This requirement still allows the Dirac fermions to have a bare mass term (vector-like under the SM), and simultaneously demands that $S_1\neq S_2$ in Fig.\ \ref{FWA}, owing to the new fermions carrying $Y\neq 0$. Consequently, two different Yukawa coupling matrices play a role in generating neutrino mass, which resolves the above-mentioned issues. With only these two scalars present in a theory, a mass flip of the vector-like fermion can not be realized for lepton AMM contributions, hence a third scalar either $S_3$ or $S_4$ must be introduced for such purpose as shown in  Fig.\ \ref{FWB}.  The presence of at least three different Yukawa couplings allows to disentangle contributions to AMM, neutrino mass and cLFV, and as mentioned above, to control the sign of the AMM contributions. 
As aforementioned,  in our setup all the BSM multiplets are assumed to be odd under $\mathcal{Z}_2$, consequently  the lightest among the neutral component fields can play the role of DM and  successfully explain the DM relic abundance. 
It is to be pointed out that the requirement of the new fermions  carrying non-zero hypercharge is an outcome of the DM-stabilizing $\mathcal{Z}_2$ symmetry, if the diagrams in Figs.\ \ref{FWA} and \ref{FWB} are supposed to exist.  This could be relaxed if a different discrete or continuous symmetry is chosen to build a model, which we do not pursue. In this work we strictly stick to $\mathcal{Z}_2$ symmetry for the fixed topology as in Fig.\ \ref{FWA} to generate neutrino mass. 
For general analyses of various topologies of neutrino mass arising by utilizing exotic vector-like fermions, see for example \cite{Babu:2001ex, deGouvea:2007qla, Angel:2012ug}.

From these Feynman diagrams and the above discussion, one sees that a common set of multiplets, either $\{F_{L,R}, S_1\}$  or $\{F_{L,R}, S_2\}$ depending on the model,  plays role in both the neutrino mass generation and in accommodating lepton AMMs data. With TeV scale vector-like fermions, the appropriate scale of neutrino masses  can be naturally  reproduced with Yukawa couplings that are comparable in order with the SM charged fermion Yukawa couplings. Furthermore, even with TeV scale vector-like fermions, the required large contributions towards both $(g-2)_e$ and $(g-2)_\mu$ can be promptly obtained via  chirality enhancement.

\begin{table*}[t!]
\centering
\resizebox{0.99\textwidth}{!}{
\begin{tabular}{|c|c|c|c|c|c|c|c|}
\hline \hline
Multiplets&Model-I&Model-II&Model-III&Model-IV&Model-V&Model-VI&Model-VII \\ \hline

$F_{L,R}$&(1,1,-1)&(1,1,-1)&\color{red}(1,2,-1/2)&\color{red}(1,2,-1/2) &(1,2,-3/2)&\color{red}(1,3,1)&\color{red}(1,3,1) \\

$S_1$&\color{red}(1,2,1/2)&\color{red}(1,2,1/2)&\color{red}(1,1,0)&\color{red}(1,1,0) &(1,1,1)&(1,2,3/2)&(1,2,3/2) \\

$S_2$&(1,2,3/2)&(1,2,3/2)&\color{red}(1,3,1)&\color{red}(1,3,1) &(1,3,-2)&\color{red}(1,2,1/2)&\color{red}(1,2,1/2) \\ \hline

$S_3$&\color{red}(1,1,0)&-&\color{red}(1,2,1/2)&- &\color{red}(1,2,1/2)&(1,3,2)&- \\

$S_4$&-&(1,1,2)&-&(1,2,3/2) &-&-&\color{red}(1,3,0)  \\ \hline \hline

\end{tabular}
}
\caption{Here we have listed only the viable models up to $SU(2)_L$ triplets that satisfy our required criteria, see text for details. By following our method,  models involving higher dimensional representations can be constructed trivially.   Multiplets containing a potential dark matter candidate are shown in red.}\label{FWtab}
\end{table*}

Having stated our criteria, the exercise is now to find a set of vector-like fermions and scalars that allow for the topologies in Figs.\ \ref{FWA} and \ref{FWB}. This leads to  
the models summarized in Table \ref{FWtab}. Here we have listed only the viable models up to $SU(2)_L$ triplets  that satisfy our above-mentioned criteria. Multiplets that contain a neutral component and thus a potential DM candidate are shown in red. By following our methodology,  models involving higher dimensional representations can be constructed trivially. It is beyond the scope of this work to study each of these models in detail. Instead, in the next section we perform a detailed analysis of the first model (Model-I) in the list.

\section{Details of Model-I}\label{sec:I}
In this section we perform a detailed analysis of Model-I. In this model, the SM particle content is extended by  three singly charged vector-like fermions $F_{L,R}$, and three scalars: a singlet and two doublets under the $SU(2)_L$ group. One of the doublets has hypercharge $1/2$ and the other has $3/2$.   As already mentioned, under the imposed $\mathcal{Z}_2$ symmetry, the SM particles are even, whereas all the BSM states are odd.  The full quantum numbers of the BSM multiplets are summarized in Table \ref{model-1}.

With this particle content, the most general  Yukawa Lagrangian consistent with all symmetries is given as 
\begin{align}
-\mathcal{L}_Y = &
y_H\overline{L}_L\ell_R H
+ y_1\overline{L}_L F_R \phi_1
+ y_2\overline{L^c}_L F_L i \tau_2 \phi_2
+ \frac{y_3}{\sqrt{2}} \overline{\ell}_R F_L \eta 
+ M_{F}\overline{F}_L F_R
+h.c. \label{L}
\end{align}
Here, $L_L$ is a left-handed lepton doublet, $\ell_R$ is a right-handed lepton, and $H$ is the SM Higgs doublet.  In the above Eq.\ \eqref{L}, for simplicity we have suppressed  generation indices. Note that the third term violates lepton number. 
Yukawa couplings of the quarks remain unchanged compared to the SM, hence we only focus on the leptonic sector. We work in the basis where the Yukawa coupling  $y_H$ and the vector-like fermion mass matrix $M_F$ are diagonal. The three new Yukawa couplings matrices $y_{1,2,3}$ are in general arbitrary. 

\begin{table}[t!]
\centering
\footnotesize
\resizebox{0.6\textwidth}{!}{
\begin{tabular}{|c|c|c|}
\hline
Multiplets& $SU(3)_C\times SU(2)_L\times U(1)_Y $ & $\mathcal{Z}_2$  \\ \hline\hline
Scalars &
\pbox{10cm}{
\vspace{2pt}
$\phi_1 (1,2,\frac{1}{2})$\\
$\phi_2 (1,2,\frac{3}{2})$\\
$\eta (1,1,0)$
\vspace{2pt}}
&\pbox{10cm}{
\vspace{2pt}
$-$\\
$-$\\
$-$
\vspace{2pt}}
\\ \hline \hline 
Vector-like fermion&
\pbox{10cm}{
\vspace{2pt}
$ F_{L,R} (1,1,-1)$
\vspace{2pt}}
&\pbox{10cm}{
\vspace{2pt}
$-$
\vspace{2pt}}
\\ \hline
\end{tabular}
}
\caption{ 
Quantum numbers of the BSM multiplets for Model-I. Both $\phi_1$ and $\eta$ contain a DM candidate.
}\label{model-1}
\end{table}

\subsection{Scalar sector}
The scalar sector of the full model consists of three neutral CP-even states $h, S^0_{1,2}$, one neutral CP-odd $A^0$, two singly charged $S^+_{1,2}$, and a doubly charged $S^{++}$. Here $h$ is identified with the SM Higgs, which does not mix with the rest of the two states $S^0_{1,2}$ due to the imposed $\mathcal{Z}_2$ symmetry. The lightest between these two states $S^0_{1,2}$ is identified as the DM.  Moreover, we assume the BSM multiplets $\eta$ and $\phi_1$ do not accrue any VEV, hence, the Goldstone bosons $G^0, G^{\pm}$ originate entirely from the SM Higgs doublet $H$. 
The  complete scalar potential for Model-I is given as 
\begin{align} \begin{aligned}\label{eq:potential}
&V = -\mu^2_H H^{\dagger}H + \sum_\varphi^{\{ \phi_1,\phi_{2}\}}\mu^2_\varphi \varphi^\dagger \varphi + \mu^2_\eta \eta^2 + (\mu_{5} H^\dagger \phi_1 \eta + h.c. ) 
     + \sum_\varphi^{\{H,\phi_1,\phi_{2}\}}\lambda_\varphi (\varphi^\dagger \varphi)^2 + \lambda_{\eta} \eta^4  
     \\ & 
     + \sum_{\varphi<\varphi'}^{\{ H,\phi_1,\phi_{2}\}}\lambda_{\varphi \varphi'} (\varphi^\dagger \varphi)( \varphi^{\prime \dagger} \varphi')
     + \sum_\varphi^{\{H,\phi_1,\phi_{2}\}}\lambda_{\varphi\eta} (\varphi^\dagger \varphi)\eta^2 
     + \sum_{\varphi<\varphi'}^{\{ H,\phi_1,\phi_{2}\}}\lambda'_{\varphi \varphi'} (\varphi^\dagger \varphi')( \varphi^{\prime \dagger} \varphi)
     \\ &
     + \left \{\lambda''_{H\phi_1}(H^{\dagger}\phi_1)^2+h.c.\right \} + \left \{ \lambda''_{\phi_1\phi_2} (H \epsilon \phi_1) (\phi_2^{\dagger}H) +h.c. \right \}.
\end{aligned}
\end{align}
We now derive the masses of the physical Higgs particles from the above potential. The mass-squared matrix $\mathcal{M}^2_{S^0}$ for the two CP-even states, written in the $\{\eta^0,\text{Re}(\phi_1^0)\}$  basis,  is 
\begin{align}
\mathcal{M}^2_{S^0}=
\begin{pmatrix}
    2\mu^2_{\eta}+\lambda_{H\eta}v^2_{H} & \mu_5 v_{H} \\
    \mu_5 v_{H} & \mu^2_{\phi_1}+\frac{(\lambda_{H\phi_1}+\lambda'_{H\phi_1}+2\lambda''_{H\phi_1})}{2}v^2_{H}
\end{pmatrix}. \label{S0}
\end{align}
The scalars that do not mix with any other fields are the SM Higgs, the CP-odd scalar, and the doubly charged scalar. The corresponding squared masses are 
\begin{align}
    &m^2_{h}=2\lambda_H v^2_{H},\\
    &m^2_{A^0}=\mu^2_{\phi_1}+\frac{(\lambda_{H\phi_1}+\lambda'_{H\phi_1}-2\lambda''_{H\phi_1})}{2}v^2_{H},\\
    &m^2_{S^{\pm\pm}}=\mu^2_{\phi_2}+\frac{\lambda_{H\phi_2}}{2}v^2_{H}, 
\end{align}
where $\phi_A$ is  $\text{Im}(\phi_1^0)$. Finally, the mass-squared matrix  for the singly charged scalars in a   basis of $(\phi_1^{\pm},\phi_2^{\pm})$ reads  
\begin{align}
\mathcal{M}^2_{S^{\pm}}=
\begin{pmatrix}
    \mu^2_{\phi_1}+\frac{\lambda_{H\phi_1}}{2}v^2_{H} & -\frac{\lambda''_{\phi_1\phi_2}}{2}v^2_{H} \\
    -\frac{\lambda''_{\phi_1\phi_2}}{2}v^2_{H} & \mu^2_{\phi_2}+\frac{(\lambda_{H\phi_2}+\lambda'_{H\phi_2})}{2}v^2_{H}
\end{pmatrix}.
\end{align}
Moreover, the mixing angle $\alpha$ ($\gamma$) between the two mixed CP-even (singly charged) states can be calculated from 
\begin{align}\label{ap}
&\tan 2\alpha= \frac{2\mu_5 v_H}{ \left(\mathcal{M}^2_{S^0}\right)_{11}-\left(\mathcal{M}^2_{S^0}\right)_{22} },
\nonumber \\
&\tan 2\gamma= \frac{\lambda''_{\phi_1\phi_2} v_H^2}{ \left(\mathcal{M}^2_{S^{\pm}}\right)_{22}-\left(\mathcal{M}^2_{S^{\pm}}\right)_{11} }. 
\end{align}
We note that the presence of non-zero $\alpha$ is crucial for generating the AMMs and for the dark matter phenomenology. Non-zero $\gamma$ is required to generate neutrino mass.  Moreover, between the two neutral physical states $S^0_{1,2}$, we will assume $S^0_{1}$ to be the ligther one and identify it as the DM candidate. Its decomposition in terms of the original fields is given by $S^0_1= \eta^0 \cos\alpha  +{\rm Re}(\phi^0_1) \sin \alpha$.  

\subsection{Lepton anomalous magnetic moments}
In the present set-up, we assume the vector-like fermions to reside  around the TeV scale. In contrast to the scotogenic case, having such a heavy mass does not require large Yukawa couplings to incorporate the $\Delta a_\ell$ data given in Eqs.\ \eqref{ae}  and \eqref{amu}. Large enough corrections to the lepton AMMs naturally arise due to a chirality flip of the vector-like fermions on the internal line, as can be seen from Fig.\ \ref{FWB}. Moreover, the sign difference for $\Delta a_e$ and $\Delta a_\mu$ is obtained by appropriately choosing the sign of the product of the Yukawa couplings that enter in this chirality enhanced  AMM term. We derive the complete NP contributions towards $(g-2)_\ell$ that is given by \cite{Leveille:1977rc}
\begin{align}
\Delta a_\ell &= \frac{m_{\ell}}{16 \pi^2} \sum_{j=1}^3 \left[
\sum_{k=1}^2 {\rm Re} \left(Y_{L,k}^{*\ell j}Y_{R,k}^{\ell j} \right) \frac{M_{F_{j}}}{M^2_{S_{k}}} G\left(\frac{M^2_{F_j}}{M^2_{S_{k}}}\right)
\right.\nonumber \\& \left.
+
 \frac{m_{\ell}^2}{4 \pi^2} \sum_{k=1}^3 \left( \left| Y_{L,k}^{\ell j} \right|^2 + \left| Y_{R,k}^{\ell j} \right|^2 \right) \frac{1}{M^2_{S_{k}}} \widetilde{G}\left(\frac{M^2_{F_j}}{M^2_{S_{k}}}\right)
+
 \frac{m_{\ell}^2}{4 \pi^2}  \left|Y_{L,4}^{\ell j}\right|^2  \frac{1}{M^2_{S_{4}}} \widetilde{G}\left(\frac{M^2_{F_j}}{M^2_{S_{4}}}\right) \right], \label{amm}  
\end{align}
where we have have defined $S_1=S^0_1$, $S_2=S^0_2$, $S_3=A^0$, and $S_4=S^{\pm\pm}$. The expressions for the loop functions are   
\begin{align}
&G\left (x\right) =  \frac{3-4x+x^2+2 \ln (x)}{(x-1)^3},  \label{gamm}
\\
&\widetilde{G}\left (x\right)=\frac{2+3x-6x^2+x^3+6x \ln x}{24(1-x)^4}.
\label{kappa}
\end{align}
The re-scaled Yukawa couplings appearing in Eq.\ \eqref{amm} are defined by 
\begin{align}
&Y_{L,1}^{\ell j} =  \frac{\sin\alpha}{\sqrt{2}}(y_1)_{\ell j}\;,\quad Y_{L,2}^{\ell j} =  \frac{\cos\alpha}{\sqrt{2}}(y_1)_{\ell j}\;.\label{yamm1}  \\ 
&Y_{R,1}^{\ell j} =  \frac{\cos\alpha}{\sqrt{2}}(y_3)_{\ell j}\;,\quad Y_{R,2}^{\ell j} =  \frac{-\sin\alpha}{\sqrt{2}}(y_3)_{\ell j}\;.\label{yamm3}
\\
&Y_{L,3}^{\ell j} =  i\;\frac{\sin\alpha}{\sqrt{2}}(y_1)_{\ell j}\;,\quad Y_{R,3}^{\ell j} =  0\;,\quad Y_{L_{\ell j}}^{\ell,4} =-(y_2)_{\ell j}\;. \label{yamm2}
\end{align}
It should be pointed out that the very first term (chirality flip term)  in Eq.\ \eqref{amm} dominates; the remaining contributions can be safely ignored for our case, which we have confirmed numerically.  In our set-up, the contribution from the SM Higgs $h$ remains unchanged, which is already  part of $a^{\rm SM}_\ell$.    
We stress here that for $\alpha=0$ the dominating first contribution to the AMMs would vanish. This can be understood from the expressions in \eqref{eq:potential} and \eqref{ap}. Vanishing $\alpha$ would correspond to vanishing $\mu_5$, and thus no triple-scalar coupling of $\phi_1$ with $\eta$ and the SM Higgs. This in turn would correspond to the absence of the AMM diagram in Fig.\ \ref{FWB}.

The off-diagonal elements in the Yukawa couplings $y_{1,3}$ will lead to  cLFV processes such as $\ell\to \ell' \gamma$. Due to the same chirality enhancement effects via the vector-like fermions, these processes impose severe constraints  on these off-diagonal Yukawa couplings. Amplitudes of these cLFV processes can be straightforwardly computed for our scenario, however, for the simplicity of our work, we assume the two Yukawa coupling matrices $y_{1,3}$ to be diagonal (meaning, small off-diagonal entries are omitted for our analysis).  
 However, non-zero but small off-diagonal entries have no impact on the results obtained in this work. There are also very stringent constraints that arise from the lepton dipole moments (for a review see Ref.\ \cite{Roberts:2010zz}) measurements for complex couplings. We avoid these constraints by demanding these $y_{1,3}$ couplings to be real.  For  completeness, here we present the generic expressions for the cLFV process $\ell\to\ell' \gamma$ for our model
 \begin{align}
&{\rm BR}\left(\ell\to\ell' \gamma\right)     = \frac{m^3_\ell \tau_\ell \alpha}{4096\pi^4}  \left(   |A^{\ell \ell'}|^2 + |A^{\ell' \ell}|^2 
\right),
\\ &
A^{\ell \ell'}= \sum_{j=1}^3 \left[  \sum_{k=1}^2 
Y^{*\ell' j}_{L,k}  Y_{R,k}^{\ell j}\;
\frac{M_{F_j}}{M^2_{S_k}} G\left(\frac{M^2_{F_j}}{M^2_{S_k}}\right) 
- \sum_{k=1}^{4}
Y^{* \ell' j}_{L,k}  Y^{\ell j}_{L,k}\; \frac{4 m_\ell}{M^2_{S_k}} 
\widetilde{G}\left(\frac{M^2_{F_j}}{M^2_{S_k}}\right)
\right],
\\ &
A^{\ell' \ell}= \sum_{j=1}^3 \left[  \sum_{k=1}^2 
Y^{*\ell j}_{L,k}  Y_{R,k}^{\ell' j}\;
\frac{M_{F_j}}{M^2_{S_k}} G\left(\frac{M^2_{F_j}}{M^2_{S_k}}\right) 
- \sum_{k=1}^{4}
Y^{* \ell j}_{R,k}  Y^{\ell' j}_{R,k}\; \frac{4 m_\ell}{M^2_{S_k}} 
\widetilde{G}\left(\frac{M^2_{F_j}}{M^2_{S_k}}\right)
\right],
 \end{align}
here $\tau_\ell$ is the lifetime of lepton $\ell$, and the Yukawa couplings and the loop functions have been defined above.

 \begin{figure*}[t!]
$$
\includegraphics[width=0.75\textwidth]{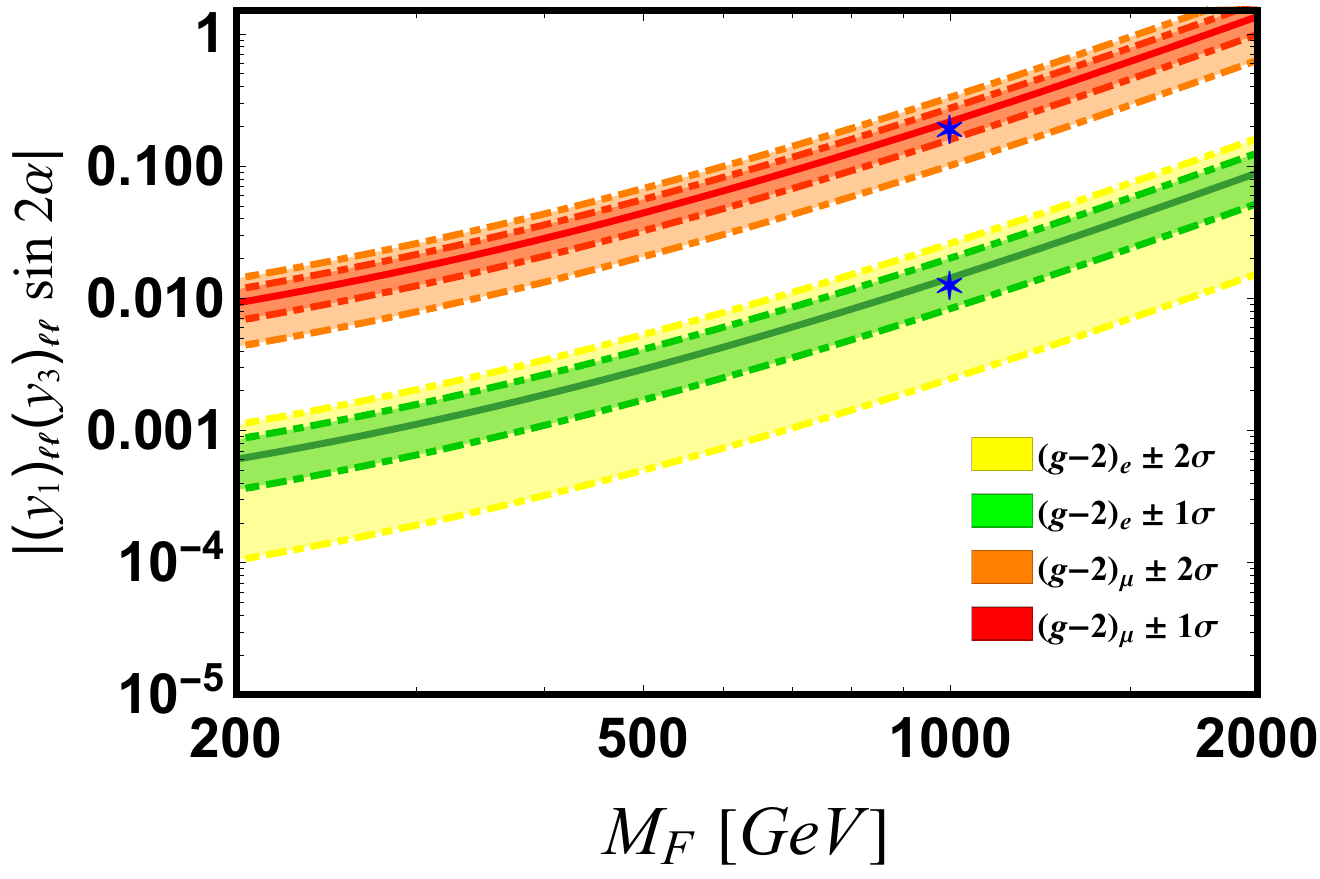}
$$
\caption{The red (green) and orange (yellow) regions indicate the experimental $1\sigma$ and $2\sigma$ bands for the muon (electron) AMM $\Delta a_{\mu}$ ($\Delta a_{e}$). The parameter space in Yukawa coupling vs.\ vector-like fermion mass plane consistent with both the electron and muon AMMs. Here we choose the mass of scalars $S_1^0$ and $S_2^0$ to be 120 GeV and 360 GeV, respectively. The blue star corresponds to the benchmark point given in Eqs.\  \eqref{bm1} -  \eqref{bm2}.  } \label{g2main}
\end{figure*}

In Fig.~\ref{g2main}, we show the parameter space  in Yukawa coupling vs.\ vector-like lepton mass plane which is consistent with  the experimentally measured values of AMMs of the electron and muon.  The red and orange regions correspond to the measured values of muon AMM within 1$\sigma$ and  2$\sigma$ allowed range respectively, whereas green and yellow regions depict the parameter space consistent with the measured value of electron AMM within 1$\sigma$ and  2$\sigma$ allowed range respectively .  For illustration purpose, here we set the mass of the scalars $S_1^0$ ($S_2^0$) to be 120 GeV (360 GeV). The blue star in Fig.~\ref{g2main} indicates to the benchmark point given in Eqs.\  \eqref{bm1} -  \eqref{bm2}.


\subsection{Neutrino mass}
The same vector-like fermions play a major role in generating radiative neutrino mass and the corresponding Feynman diagram is presented in Fig. \ref{FWA}.  The loop is completed via the propagation of the singly charged scalars $S^+_{1,2}$, and we obtain the following expression for the neutrino mass matrix for Model-I 
\begin{align}
\mathcal{M}^{\nu}_{ij}=
\frac{\sin 2\gamma}{16\pi^2} &\sum_{\alpha=1}^3
\left[ (y_1)_{i\alpha}(y_2)_{j\alpha} + (y_1)_{j\alpha}(y_2)_{i\alpha} \right]
M_{F_\alpha}
\left[   
\frac{M^2_{S^+_1}\ln{\frac{M^2_{S^+_1}}{M^2_{F_\alpha}}}}{M^2_{S^+_1}-M^2_{F_\alpha}}  -\frac{M^2_{S^+_2}\ln{\frac{M^2_{S^+_2}}{M^2_{F_\alpha}}}}{M^2_{S^+_2}-M^2_{F_\alpha}}  
\right]. \label{nu}
\end{align}
Here the mixing angle $\gamma$ between the singly charged physical particles is defined in 
Eq.\ \eqref{ap}.   
We stress here that for $\gamma=0$ the neutrino mass would vanish. This can be understood from the expressions in \eqref{eq:potential} and \eqref{ap}. Vanishing $\gamma$ would correspond to vanishing $\lambda_{\phi_1 \phi_2}''$, and thus no quartic scalar coupling of $\phi_1$ with $\phi_2$ and a SM Higgs pair. This in turn would correspond to the absence of the neutrino mass diagram in Fig.\ \ref{FWA}. 

Note that to reproduce correct lepton mixing, one must have a non-trivial structure for the Yukawa coupling matrix $y_2$, since $y_1$ is taken to be diagonal. This however, does not conflict with lepton flavor violating $\ell\to \ell' \gamma$ processes mediated by the doubly charged scalars, since $(y_2)_{ij}\sim 10^{-5}$ in order to generate the correct neutrino mass scale.  
In the next subsection, we provide a realistic fit to neutrino mass spectrum.

\subsection{A combined fit to data}
To demonstrate the viability of our proposed framework, here we present a combined fit to reproduce the experimental results.  The expressions for $(g-2)_e$ and $(g-2)_\mu$ are given in Eq.\ \eqref{amm}, and their corresponding measured values can be found in Eqs.\ \eqref{ae} and \eqref{amu}. Furthermore, from the neutrino mass formula Eq.\ \eqref{nu}, one needs to successfully incorporate two mass-squared differences, three mixing angles, and one Dirac CP phase. The associated measured values in the experiments are summarized in Table \ref{nudata}.  The neutrino mass matrix can be parameterized as follows
\begin{align}
&\mathcal{M}_\nu=U_{{\rm PMNS}}\; {\rm diag}\{m_1,m_2,m_3\}\;    U_{{\rm PMNS}}^T\, ,
\\
&U_{{\rm PMNS}}=\begin{pmatrix}
c_{12}c_{13}&s_{12}c_{13}&s_{13} e^{-i\delta}\\
-s_{12}c_{23}-c_{12}s_{23}s_{13}e^{i\delta}&c_{12}c_{23}-s_{12}s_{23}s_{13}e^{i\delta}&s_{23}c_{13}\\
s_{12}s_{23}-c_{12}c_{23}s_{13}e^{i\delta}&-c_{12}s_{23}-s_{12}c_{23}s_{13}e^{i\delta}&c_{23}c_{13}
\end{pmatrix} \begin{pmatrix}
1&0&0\\
0&e^{i\frac{\alpha_{21}}{2}}&0\\
0&0&e^{i\frac{\alpha_{31}}{2}}
\end{pmatrix},
\end{align}
where $m_i$ are real eigenvalues and we have defined $c_{ij}= \cos \theta_{ij},  s_{ij}= \sin \theta_{ij}$. In the PMNS mixing matrix there exist three physical phases, one Dirac phase  $\delta\equiv \delta_{CP}$ and two Majorana phases $\alpha_{21,31}$, where we have used the particle data group (PDG) parametrization. In this work, we assume a normal  ordering for neutrino masses that corresponds to $m_1 < m_2 < m_3$, which still is favored by  
 oscillation data \cite{deSalas:2020pgw,1809173}. 

\color{black}

\begin{table}[t!]
\centering
\footnotesize
\resizebox{0.88\textwidth}{!}{
\begin{tabular}{|c|c||c|c|}\hline
Parameter& Best fit $\pm 1 \sigma$&Parameter& Best fit $\pm 1 \sigma$\\ [1ex] \hline

$\Delta m_{21}^2\;(10^{-5} \, {\rm eV}^2)$& $7.50_{-0.20}^{+0.22}$ &$\sin^2\theta_{12}$ &$0.318\pm 0.016$
\\ \hline

$\Delta m_{31}^2\; (10^{-3} \, {\rm eV}^2)$& $2.56_{-0.04}^{+0.03}$
&$\sin^2\theta_{23}$  &$0.566_{-0.022}^{+0.016}$
\\ \hline

$\delta_{CP}$& $1.20_{-0.14}^{+0.23}\pi$
&$\sin^2\theta_{13}$  &$0.02225_{-0.00078}^{+0.00055}$
\\ \hline

\end{tabular}
}
\caption{ 
Current experimental values of the neutrino observables with their corresponding $1\sigma$ uncertainties taken from Ref.\ \cite{deSalas:2020pgw}. 
}\label{nudata}
\end{table}
With all these in hand, we perform a combined numerical analysis and provide a benchmark point in the following 
\begin{align}
&M_{F_\alpha}=1\, {\rm TeV};\;\; M_{S^0_1}=0.12\, {\rm  TeV},\;\; M_{S^0_2}=3 M_{S^0_1}; \;\; M_{S^+_1}=0.46 \, {\rm TeV};\,\;\; M_{S^+_2}= 3 M_{S_1^+},  \label{bm1}
\\
&\sin \alpha= 0.1,\;\;\lambda^{''}_{\phi_1\phi_2}=8.325\; \times 10^{-6}\;,
\\
&y_1=\begin{pmatrix}
0.3662&0&0\\
0&-1.0141&0\\
0&0&-0.43913
\end{pmatrix},\;\;\; y_3=\begin{pmatrix}
-0.19428&0&0\\
0&-1.07051&0\\
0&0&0.14602
\end{pmatrix},\label{bmy13}
\\
&y_2=10^{-5}\begin{pmatrix}
-0.10116+0.07932\,i& -1.01099+0.62701\,i&-0.49019-0.30200\,i\\
-0.03471-0.04519\,i&  0.31733-0.69407\,i&
0.98401+0.8099\,i\\
0.53005-0.9659\,i& -0.05210+0.67329\,i&
-1.00145+0.82294
\end{pmatrix}.\label{bm2}
\end{align} \color{black}
Yukawa couplings of order $y_2\sim 10^{-5}$ automatically satisfy all experimental constraints, including  cLFV processes. 
The  values of the theory parameters corresponding to this benchmark point  successfully reproduce all the observables both in the neutrino sector as well as AMMs of the electron and the muon, we list the predictions in Table \ref{fit}. Since an explanation of the lepton AMMs demands Yukawa couplings of order unity (as can be seen from Eq.\ \eqref{bmy13}), and the same Yukawa couplings enter in neutrino mass generation, it can be easily understood that  $y_2\;\lambda^{''}_{\phi_1\phi_2}\sim 10^{-10}$ (instead of $y_2^2\sim 10^{-10}$ as in Eq.\ \eqref{quntifyNU}) must be satisfied to reproduce the correct neutrino mass scale. 
Regarding the smallness of $\lambda^{''}_{\phi_1\phi_2}$, 
we recall that it is the coefficient of the  quartic coupling $(H \epsilon \phi_1) (\phi_2^{\dagger}H)$ responsible for mixing the two singly charged states, as defined in Eq.\ \eqref{ap}. In the limit of $\lambda^{''}_{\phi_1\phi_2}\to 0$ neutrino masses are zero, because the theory regains the accidental lepton number conservation of the SM.  The chosen DM ($S^0_1$)  mass of 120 GeV and the associated mixing angle $\sin \alpha$ for this benchmark point will be shown to be consistent with both DM detection bounds as well as  DM  relic abundance as detailed in the next section. 
It should be pointed out that in the DM analysis more parameters such as the Higgs portal coupling and lepton coupling portal play role, which are not fixed by the fit performed above.  Moreover, the mass of the doubly charged scalar is not determined from the fit, which for simplicity, we choose  to be degenerate in mass  with its singly charged partner to be consistent with $T$ parameter constraints. However a splitting of order $\mathcal{O}(100)$ GeV is still allowed \cite{Kanemura:2012rs}. 

\begin{table*}[th!]
\begin{minipage}{.65\linewidth}
\centering
\begin{tabular}{|c|c|}\hline
Quantity  & Fit value   \\ [1ex] \hline
$\Delta a_e$&$-8.696\times 10^{-13}$\\ 
$\Delta a_\mu$&$2.744\times 10^{-9}$\\ \hline \hline
$\Delta m^2_{21}\;(10^{-5}\, {\rm eV}^2)$&7.525\\ 
$\Delta m^2_{31}\;(10^{-3} \, {\rm eV}^2)$&2.552\\  \hline\hline
$\sin^2 \theta_{12}$&0.3171\\ 
$\sin^2 \theta_{23}$&0.5638\\ 
$\sin^2 \theta_{13}$&0.02216\\ \hline \hline
$\delta_{CP}$&223.8$^{\circ}$\\  \hline
\end{tabular}
\end{minipage} \hspace{-2.5cm}
\begin{minipage}{.5\linewidth}
\centering
\begin{tabular}{|c|c|}\hline
Quantity  & Fit value   \\ [1ex] \hline
$m_1\;{\rm (eV)}$&0.00812\\ 
$m_2\;{\rm (eV)}$&0.01188\\ 
$m_3\;{\rm (eV)}$&0.05117\\ \hline \hline
$m_{\rm cos}\;{\rm (eV)}$&0.07118\\ 
$m_{\beta}\;{\rm (eV)}$&0.01207\\ 
$m_{\beta\beta}\;{\rm (eV)}$&0.00167\\ \hline \hline
$\alpha_{21}$&188.8$^{\circ}$\\  \hline
$\alpha_{31}$&311.9$^{\circ}$\\  \hline
\end{tabular}
\end{minipage}
\caption{  Fit values of some of the observables for our benchmark points given in Eqs.\  \eqref{bm1} -  \eqref{bm2}. Here $m_{\rm cos}=\sum_{i} m_{i}$, $m_{\beta}=\sqrt{\sum_{i} |U_{e i}|^{2} m_{i}^2}$ is the effective mass parameter for beta decay, and $m_{\beta \beta}= | \sum_{i} U_{e i}^{2} m_{i} |$ is the effective mass parameter for neutrinoless double beta decay. }
\label{fit}
\end{table*}


\subsection{Dark matter phenomenology}\label{SEC-4}
In this subsection, we analyze the Dark Matter (DM) phenomenology in Model-I, where lepton anomalous magnetic moments, neutrino masses and mixings are successfully generated. As aforementioned, in this model the presence of a discrete symmetry $\mathcal{Z}_2$ stabilizes the DM particle. The  newly introduced scalars ($\phi_1, \phi_2$ and $\eta$) and vector-like leptons ${F_{L,R}}$ are odd under this discrete symmetry, whereas the SM particles are even.  The lightest neutral particle among the new ones qualifies to be a DM candidate. In our setup for Model-I, the dark matter candidate will be an admixture of neutral components of the doublet $\phi_1$ and the singlet $\eta$. As one can see from Eq.~\eqref{amm}, one needs to introduce mixing between these two fields to successfully address electron and muon $g-2$ anomalies.  
Hence, the dark matter can be neither pure singlet type \cite{Steele:2013fka,Cline:2013gha,Guo:2010hq,Bandyopadhyay:2010cc,Burgess:2000yq,McDonald:1993ex,Yaguna:2011qn,Belanger:2012zr,Drozd:2014yla,Feng:2014vea,Bhattacharya:2016qsg,Bhattacharya:2016ysw,Arcadi:2019lka} nor pure inert doublet type \cite{Ma:2006km,Barbieri:2006dq,Gustafsson:2007pc,Agrawal:2008xz,Nezri:2009jd,Arina:2009um,Gong:2012ri,LopezHonorez:2006gr,Goudelis:2013uca}. Rather, it will be singlet-doublet scalar dark matter \cite{Cohen:2011ec}. While scalar singlet dark matter is tightly constrained from DM direct detection experiments \cite{Feng:2014vea,Bhattacharya:2016ysw,Aprile:2018dbl,Akerib:2016vxi,Cui:2017nnn}, inclusion of mixing with an additional doublet can introduce new additional interactions producing the right amount of relic density, and which  can potentially allow for evasion of direct detection bounds \cite{Aprile:2018dbl,Akerib:2016vxi,Cui:2017nnn} for a large region of parameter space. 

\begin{figure}[t!]
\centering
$$
\includegraphics[scale=0.5]{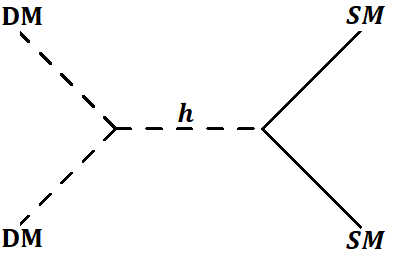}\hspace{0.55in}
\includegraphics[scale=0.5]{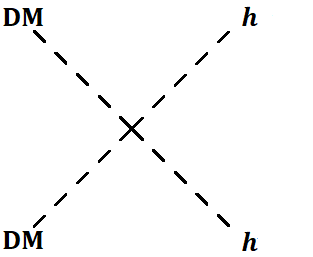}\hspace{0.55in}
\includegraphics[scale=0.5]{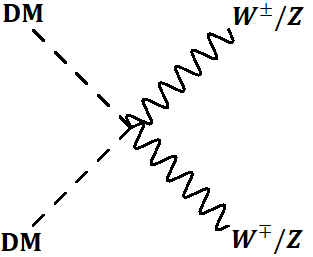}
$$
$$
\includegraphics[scale=0.5]{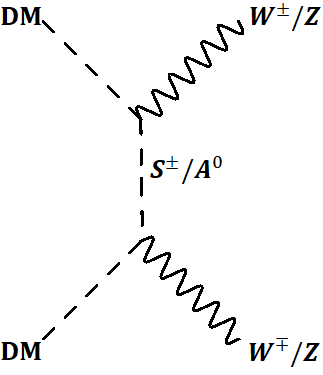}\hspace{0.58in}
\includegraphics[scale=0.5]{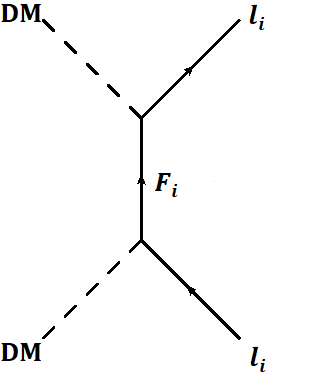}
$$
\caption{Relevant Feynman diagrams that contribute to the annihilation of the DM. 
}\label{DM_ann}
\end{figure}

The dominant processes  that contribute to the annihilation of the DM  particle are shown in Fig.~\ref{DM_ann}.
\begin{figure*}[t!]
$$
\includegraphics[width=0.48\textwidth]{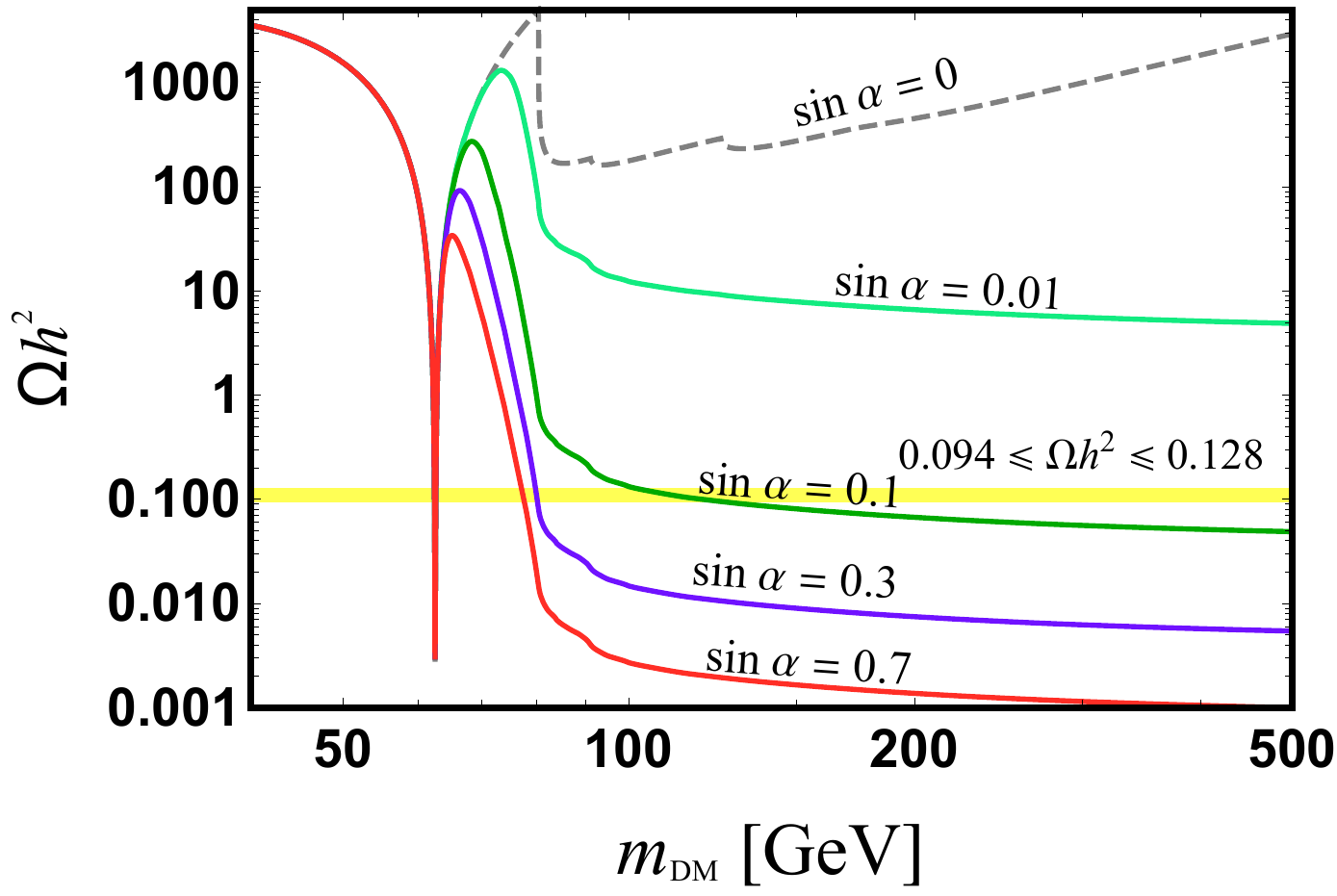}
\includegraphics[width=0.48\textwidth]{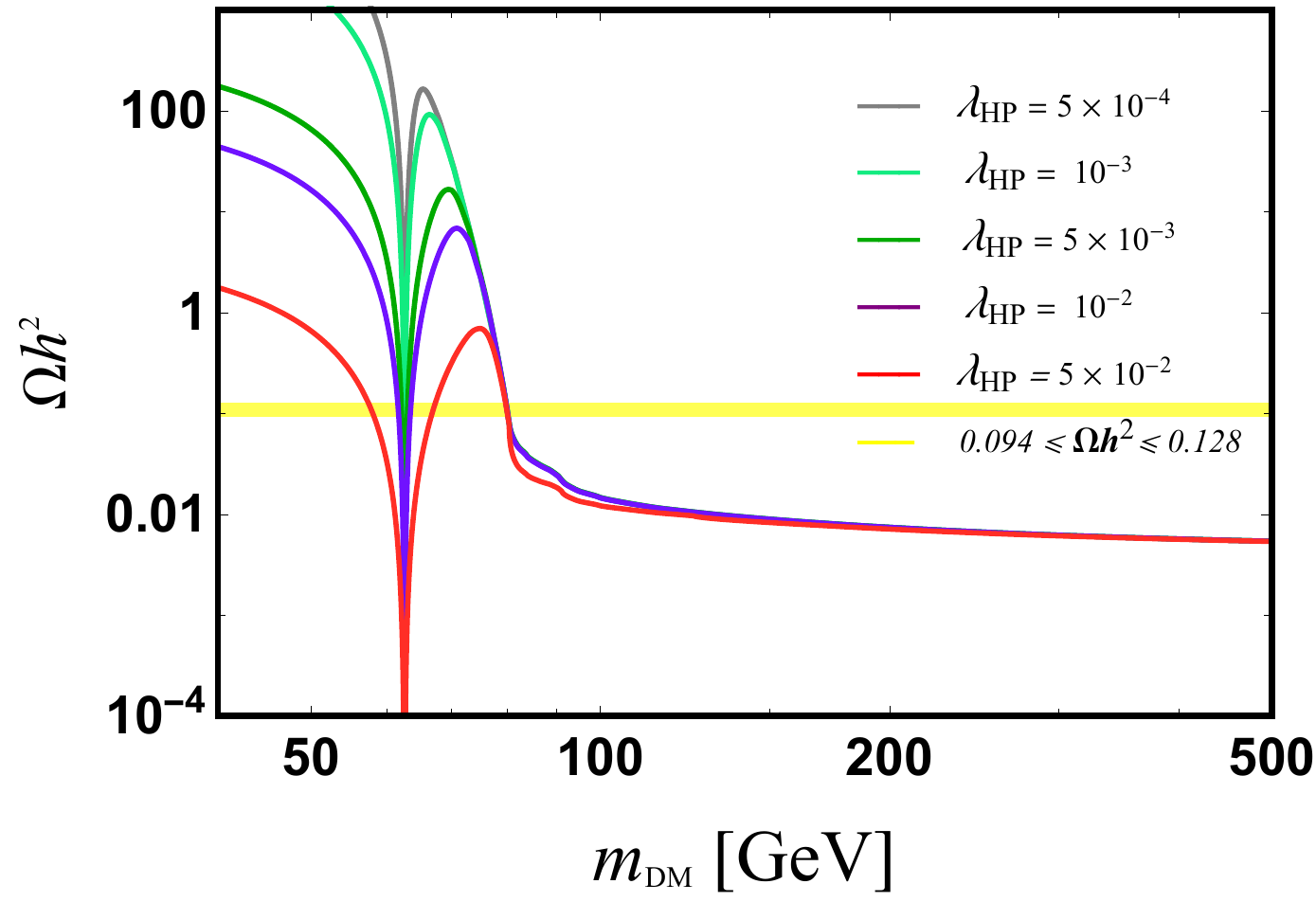}
$$
$$
\includegraphics[width=0.48\textwidth]{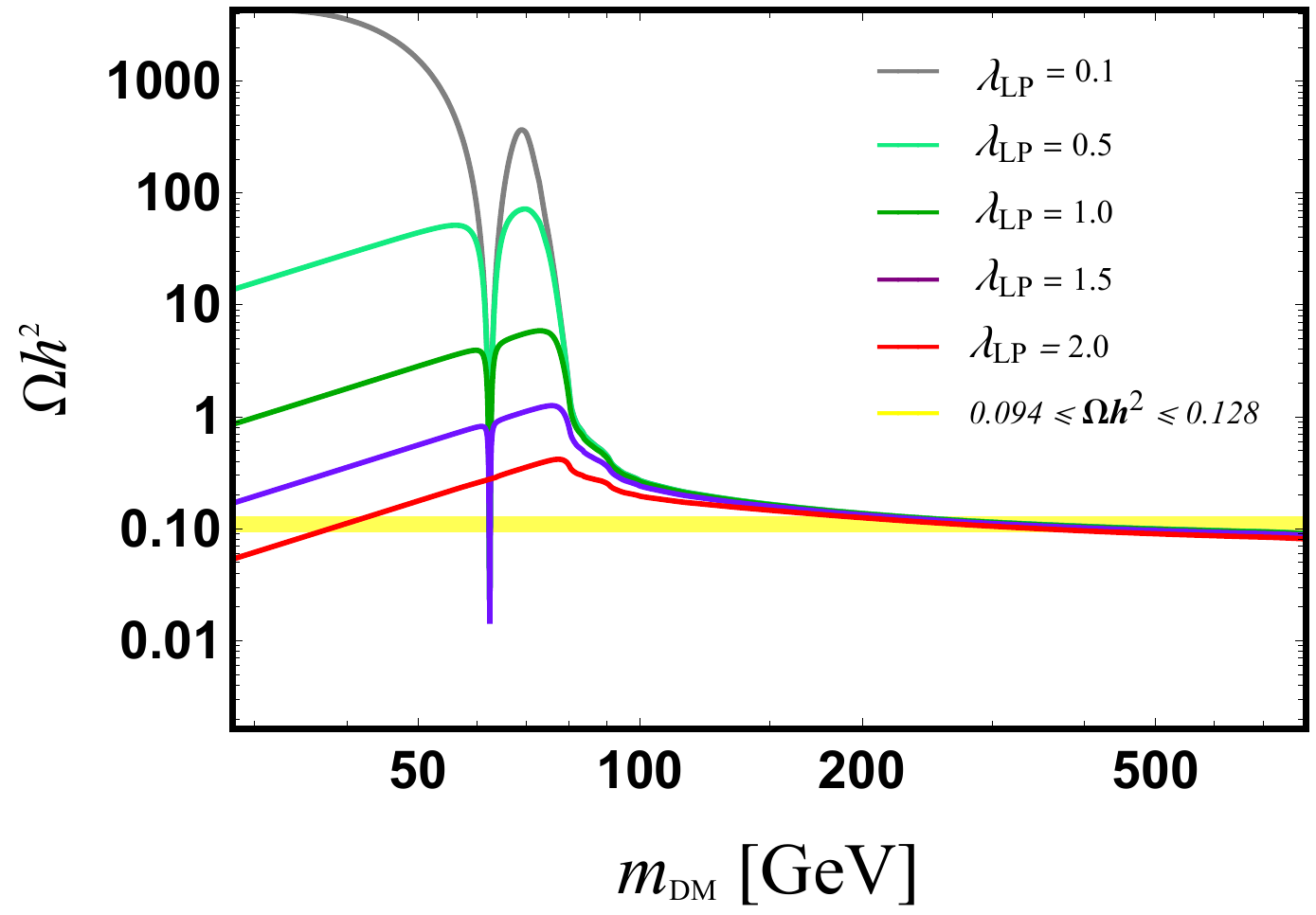}
$$
\caption{DM relic density ($\Omega h^2$) as a function of DM mass ($m_{\rm DM}$). {\it Top left:} for different choices of mixing angle: $\sin \alpha=0.0$ (grey-dotted), $\sin \alpha=0.01$ (light-green), $\sin \alpha=0.1$ (green), $\sin \alpha=0.3$ (violet) and $\sin \alpha=0.7$ (red). Here we choose the Higgs-portal coupling $\lambda_{\rm HP}=10^{-3}$ and the leptonic portal coupling $\lambda_{\rm LP}=0.1$ for illustration. {\it Top right:} for various choices of the Higgs-portal coupling: $\lambda_{\rm HP}=5\times10^{-2}$ (red), $\lambda_{\rm HP}=10^{-2}$ (violet), $\lambda_{\rm HP}=5\times10^{-3}$ (green), $\lambda_{\rm HP}=10^{-3}$ (light-green), and $\lambda_{\rm HP}=5\times10^{-4}$ (grey). The mixing angle $\sin \alpha=0.3$ and the leptonic portal coupling $\lambda_{\rm LP}=0.1$ are chosen for illustration. {\it Bottom:} for different choices of leptonic portal coupling: $\lambda_{\rm LP}=0.1$ (grey), $\lambda_{\rm LP}=0.5$ (light-green), $\lambda_{\rm LP}=1.0$ (green), $\lambda_{\rm LP}=1.5$ (violet), and $\lambda_{\rm LP}=2.0$ (red). Here we choose the Higgs-portal coupling $\lambda_{\rm HP}=10^{-3}$ and the mixing angle $\sin \alpha=0.07$ for illustration. The yellow band indicates the WMAP-observed relic density bound \cite{Hinshaw_2013}. For all the panels, we set the vector-like lepton  mass to be 1 TeV.
} \label{relic12}
\end{figure*}
In our case, the DM can annihilate to  SM particles through $s$-channel Higgs-mediated processes (Higgs-portal). These Higss-portal processes can be particularly important when the DM mass is close to half of the Higgs boson mass. Above this mass regime, the DM annihilation to gauge bosons 
(possible because it is partly a doublet) contributes  dominantly to the annihilation processes. In this mass region, the DM annihilation through the $t$-channel exchange is usually smaller than the contribution from the 4-point vertex shown in Fig.~\ref{DM_ann}. Finally, the presence of the vector-like leptons opens up new annihilation modes for DM via the $t$-channel processes (lepton-portal) as shown in Fig.~\ref{DM_ann}. In the low mass region, these leptonic portal processes become significant in addition to the Higgs-portal processes.  For the DM analysis, we will denote the lepton portal coupling by $\lambda_{\rm LP}$. 
Since the DM is identified to be the state $S^0_1$, its lepton portal couplings with $F_{L_j}$ ($F_{R_j}$) is $\sqrt{2}Y^{\ell j}_{L,1}$ ($\sqrt{2}Y^{\ell j}_{R,1}$) which can be read off from Eqs.\ \eqref{yamm1} and \eqref{yamm3}. On the other hand, $\lambda_{\rm HP}$ represents the Higgs portal coupling, which is defined by  $\lambda_{\rm HP}=  (\lambda_{H\phi_1}+\lambda^{'}_{H\phi_1}+2\lambda^{''}_{H\phi_1})\sin^2\alpha\;+2 \lambda_{H\eta}\;\cos^2\alpha$. That is, our DM particle $S_1^0$ couples via 
\begin{equation}
    {\cal L} = \frac{\lambda_{\rm HP}}{2} \, (S_1^0)^2 \, H^\dagger H + 
    \frac{\lambda_{\rm LP}}{\sqrt{2}}  \, S_1^0\; \overline{\ell}_{L,R} \, F_{R,L} 
    \,.
\end{equation}

\begin{figure*}[t!]
$$
\includegraphics[width=0.5\textwidth]{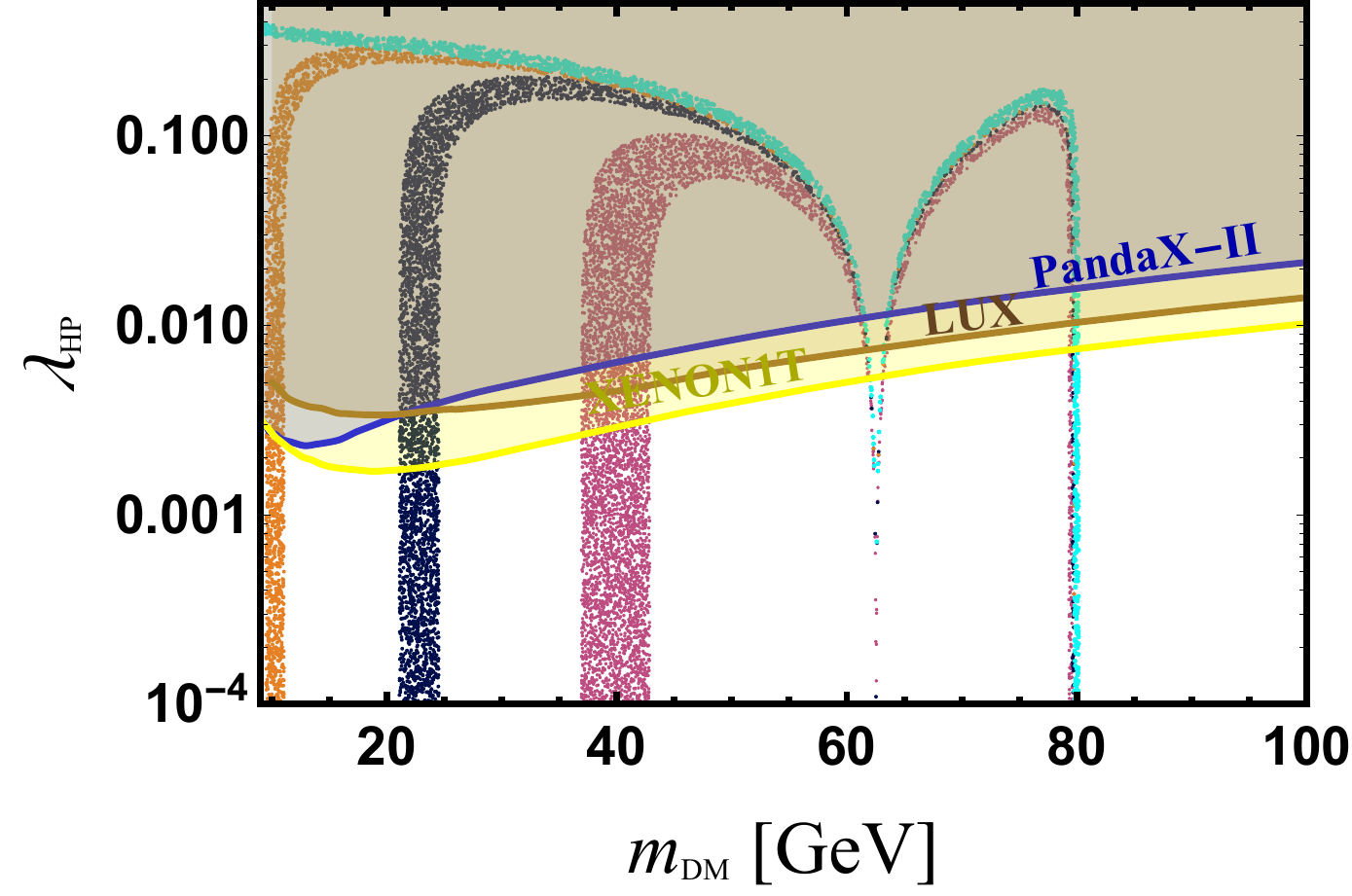}
\includegraphics[width=0.5\textwidth]{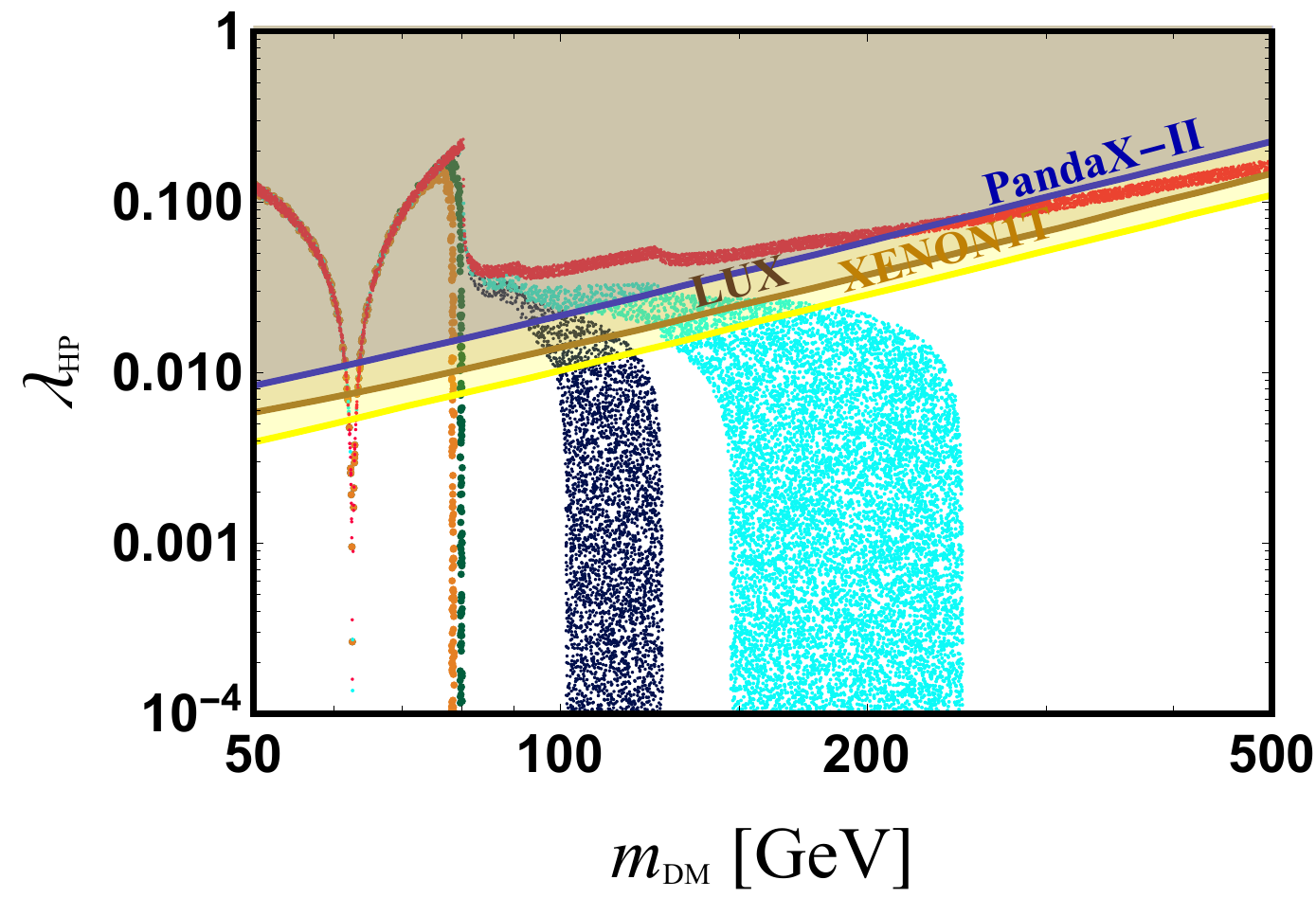}
$$
\caption{The parameter space in Higgs-portal coupling ($\lambda_{\rm HP}$) vs.\ DM mass ($m_{\rm DM}$) plane consistent with the WMAP relic density constraint. {\it Left:} regions are shown for different choices of leptonic portal coupling: $\lambda_{\rm LP}=0.5$ (cyan), $\lambda_{\rm LP}=1.0$ (orange), $\lambda_{\rm LP}=1.5$ (blue), and $\lambda_{\rm LP}=2.0$ (pink). Here the mixing angle $\sin \alpha =0.3$ is kept fixed.
{\it Right:} regions are shown for different choices of mixing angle: $\sin\alpha=0.01$ (red), $\sin\alpha=0.08$ (cyan), $\sin\alpha=0.1$ (blue), $\sin\alpha=0.3$ (green), and $\sin\alpha=0.5$ (orange). Here we choose the leptonic portal coupling $\lambda_{\rm LP}=0.1$. The color shaded regions with solid boundary line denote the excluded parameter space by various current direct detection experiments: brown region from LUX-2017 \cite{Akerib:2016vxi}; blue region from PandaX-II   \cite{Wang:2020coa}; yellow region from XENON1T (2018) \cite{Aprile:2018dbl}. We set the vector-like lepton  mass to be 1 TeV.
} \label{relic4}
\end{figure*}

Let us quantify the DM phenomenology further. For our DM analysis, we have inserted our model in {\tt{micrOMEGAs}} \cite{Belanger:2014vza,Belanger:2018ccd} and scan over the parameter space to analyze relic abundance and direct detection constraints. For the rest of the analysis we fix the vector-like lepton mass to be 1 TeV. As mentioned in the above paragraph, in our case the viable DM mass range which is consistent with the WMAP relic density constraint can be divided into three regions. In the low mass regime ($m_{\rm DM} \lesssim 55 $ GeV), the main annihilation channel of DM  is via the leptonic $t$-channel processes, mediated by the vector-like leptons. Since the $s$-wave and $p$-wave contributions of this leptonic channels are helicity-suppressed \cite{Gaviria:2020uio,Berlin:2014tja}, the $d$-wave contribution becomes dominant for the case of DM annihilation into electron-positron and muon-antimuon pairs. For DM annihilation into tau leptons, the $s$-wave and $p$-wave contributions become dominant compared to the $d$-wave contribution \cite{Gaviria:2020uio}. In  Fig.~\ref{relic12} (bottom), we analyze the DM relic density as a function of DM mass for various leptonic portal couplings ($\lambda_{\rm LP}$). Here we set the Higgs portal coupling $\lambda_{\rm HP}=10^{-3}$ and the mixing angle  $\sin \alpha=0.07$ for  illustration. For simplicity, we also choose the leptonic portal coupling to be same for all the three leptons. For illustrating this further, we have also scanned the parameter space in Higgs-portal coupling ($\lambda_{\rm HP}$) vs.\ DM mass ($m_{\rm DM}$) plane consistent with the WMAP relic density constraint for different choices of leptonic portal coupling in Fig.~\ref{relic4} (left). In the intermediate mass region (55 GeV $\lesssim m_{\rm DM} \lesssim 75 $ GeV), the dominant contribution to the DM annihilation comes from the $s$-channel Higgs mediated process. In Fig.~\ref{relic12} (top right), we analyze the DM relic density as a function of DM mass for various Higgs-portal couplings ($\lambda_{\rm HP}$). The mixing angle $\sin \alpha=0.3$ and  leptonic portal coupling $\lambda_{\rm LP}=0.1$ are chosen for better illustration. In the high mass regime ($m_{\rm DM} \gtrsim$ 75 GeV), the relic density of DM depends on the mixing angle. In this parameter space, the dominant contribution to the DM annihilation cross-section comes from the weak gauge bosons channels. In Fig.~\ref{relic12} (top left), we show the effect of varying the mixing angle on the relic density of DM for a fixed value of $\lambda_{\rm HP}=10^{-3}$ and $\lambda_{\rm LP}=0.1$. As the mixing angle increases, the annihilation cross-section of DM into weak gauge bosons also increases. Due to this, the  WMAP relic density constraint for DM can be satisfied for higher DM masses as well. For illustrating this, we also show the parameter space in Higgs-portal coupling ($\lambda_{\rm HP}$) vs.\ DM mass ($m_{\rm DM}$) plane consistent with the WMAP relic density constraint for various choices of mixing angle in Fig.~\ref{relic4} (right).\\ 

\begin{figure*}[t!]
$$
\includegraphics[width=0.5\textwidth]{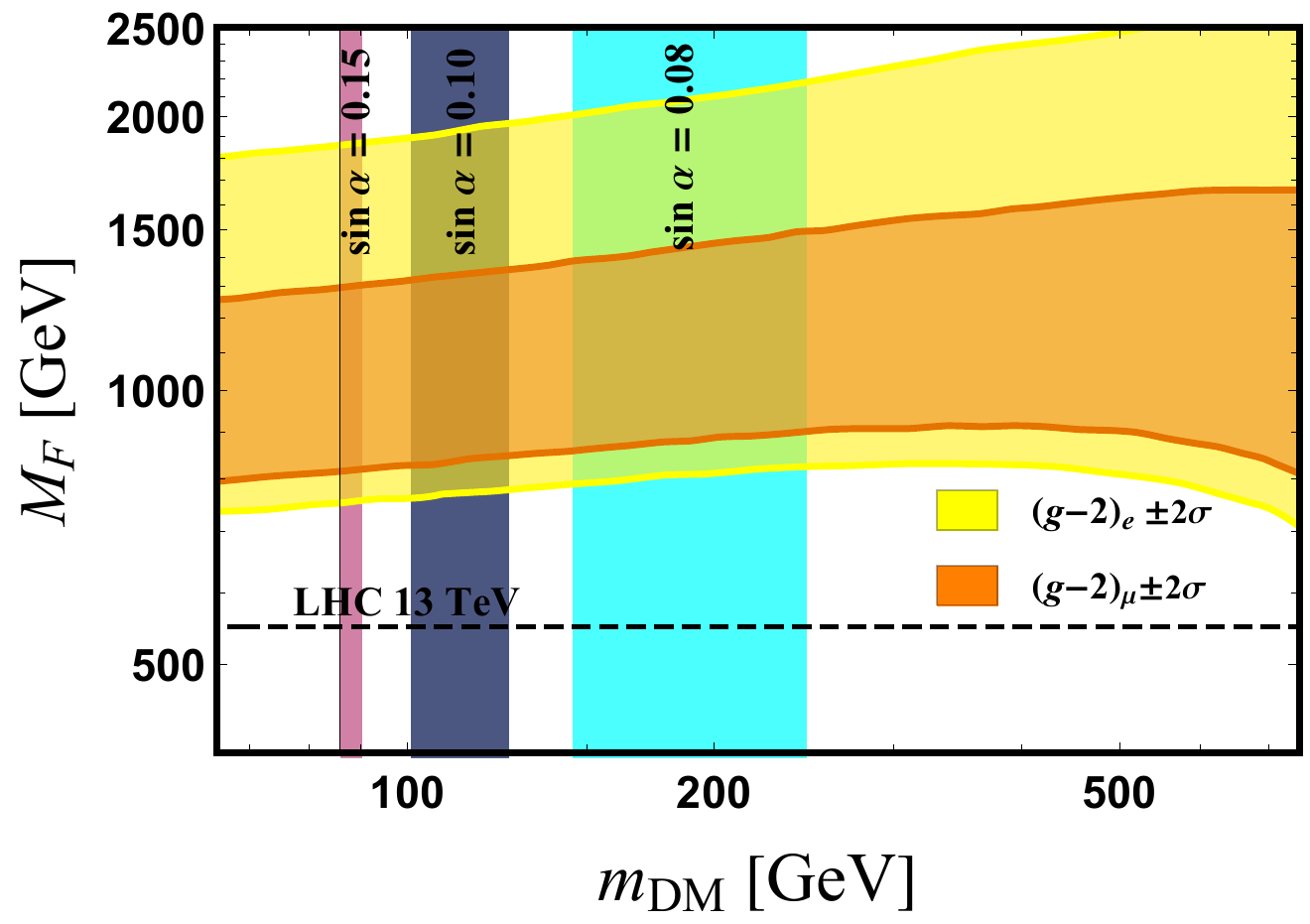}
\includegraphics[width=0.5\textwidth]{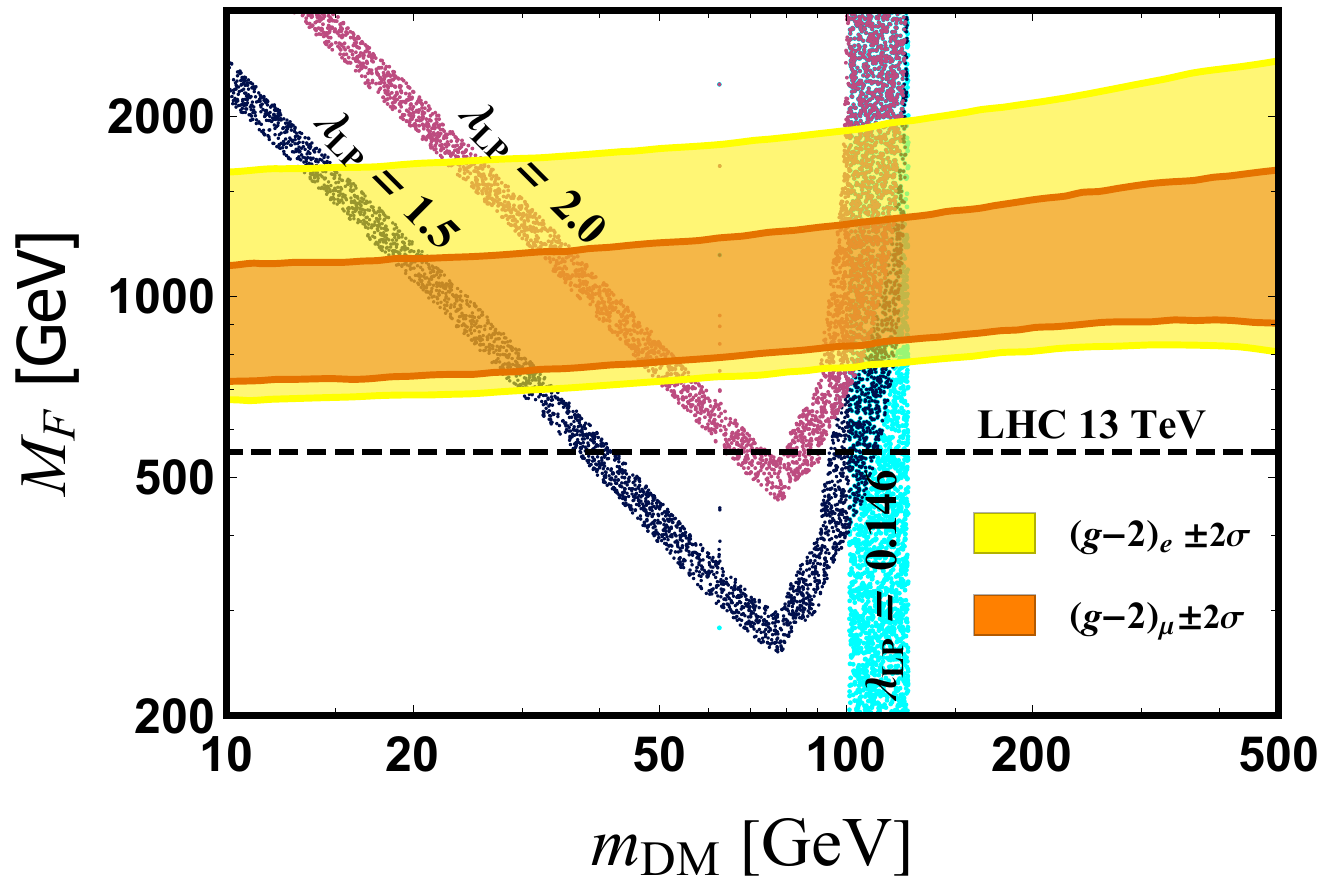}
$$
\caption{The parameter space in vector-like lepton mass ($M_F$) vs.\ DM mass ($m_{\rm DM}$) plane consistent with both the electron and muon AMMs. The orange (yellow) region indicates the experimental $2\sigma$ band for the muon (electron) AMM $\Delta a_{\mu}$ ($\Delta a_{e}$). {\it Left:} the colored vertical bands represent
the regions that are consistent with the WMAP relic density constraint for different choices of mixing angle: $\sin \alpha=0.08$ (cyan), $\sin \alpha=0.1$ (blue), and $\sin \alpha=0.15$ (pink). Here we fix the Higgs-portal coupling $\lambda_{\rm HP}$ to be $10^{-3}$. {\it Right:} the colored regions are shown for different choices of leptonic portal coupling that are consistent with the WMAP relic density constraint: $\lambda_{\rm LP}=0.146$ (cyan), $\lambda_{\rm LP}=1.5$ (blue), and $\lambda_{\rm LP}=2.0$ (pink). Here we fix the mixing angle $\sin \alpha=0.1$ and the Higgs-portal coupling $\lambda_{\rm  HP}=4\times 10^{-4}$. For both the panels, we fix the product $(y_1)_{\ell \ell}(y_3)_{\ell \ell} \sin 2\alpha $ to be same as the benchmark value given in Eqs.\  \eqref{bm1} -  \eqref{bm2}. The horizontal dashed-line indicates the bound on the vector-like lepton mass from the 13 TeV LHC data \cite{Khachatryan:2016sfv}.   } \label{g2maindm}
\end{figure*}

In addition to the DM relic density study, we also consider the constraints from various DM direct detection experiments. In our model, the DM can interact with nuclei dominantly via  $t$-channel Higgs boson exchange. The corresponding spin independent DM-nucleon scattering cross-section is estimated in \cite{Cline:2013gha,Bhattacharya:2016ysw}. Using this,  we recast the limits from LUX-2017 \cite{Akerib:2016vxi}, PandaX-II \cite{Wang:2020coa} and XENON1T (2018) \cite{Aprile:2018dbl} experiments for our model, which are shown as brown, blue   and yellow region, respectively, in Fig.~\ref{relic4}. As one can see, we can satisfy all the present bounds from DM direct detection experiments for a large region of the parameter space.

 Finally, in Fig.\ \ref{g2maindm},
 we show the correlation between the vector-like lepton mass and DM mass in order to obtain the correct experimental values of $(g-2)_{e,\mu}$ as well as the DM relic abundance.   The orange and yellow region  in vector-like lepton mass ($M_F$) vs.\ DM mass ($m_{\rm DM}$) plane depict the parameter space which can address electron and muon $g-2$ anomalies respectively. The pink, blue and cyan shaded bands represent the parameter space consistent with the DM relic abundance ($0.094 \leq \Omega h^2 \leq 0.128$). To illustrate the correlation, in left panel of  Fig.\ \ref{g2maindm}, we set the Higgs-portal coupling $\lambda_{\rm HP}$ to be $10^{-3}$ and vary the mixing angle: $\sin \alpha=0.08$ (cyan), $\sin \alpha=0.1$ (blue), and $\sin \alpha=0.15$ (pink). On the other hand, in right panel of  Fig.\ \ref{g2maindm}, the colored regions are shown for different choices of leptonic portal coupling that are consistent with the WMAP relic density constraint: $\lambda_{\rm LP}=0.146$ (cyan), $\lambda_{\rm LP}=1.5$ (blue), and $\lambda_{\rm LP}=2.0$ (pink), while  fixing the mixing angle $\sin \alpha=0.1$ and the Higgs-portal coupling $\lambda_{\rm  HP}=4\times 10^{-4}$. For both the panels, we fix the product $(y_1)_{\ell \ell}(y_3)_{\ell \ell} \sin 2\alpha $ to be same as the benchmark value given in Eqs.\  \eqref{bm1} -  \eqref{bm2}. The horizontal dashed-line indicates the bound on the vector-like lepton mass from the 13 TeV LHC data \cite{Khachatryan:2016sfv}, cf.\ Section \ref{SEC-5}. As we can see  from Fig.\ \ref{g2maindm},  there is a significant region of parameter space  (intersection zones) which can accommodate the correct experimental values of $(g-2)_{e,\mu}$ as well as the DM relic abundance.

\subsection{Collider implications}\label{SEC-5}
\begin{figure*}[t!]
\includegraphics[width=0.48\textwidth]{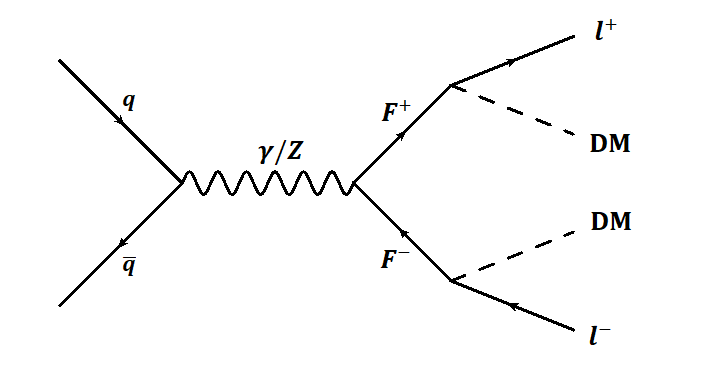}
\includegraphics[width=0.48\textwidth]{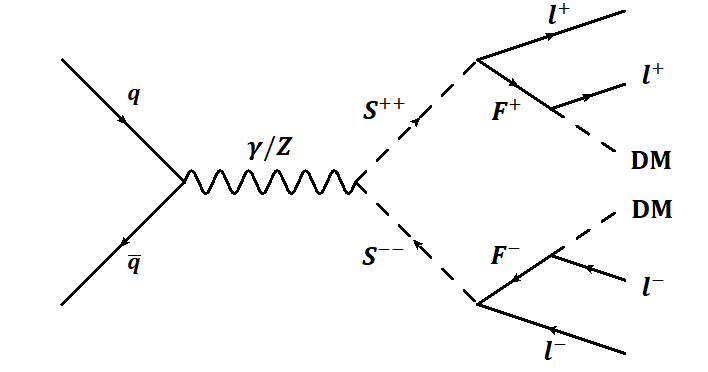}
\caption{Feynman diagrams for the collider signal of DM at the LHC. 
} \label{fyncoll}
\end{figure*}
Here we discuss the collider phenomenology associated with the dark matter in our model. Especially, the presence of  doubly charged scalar $S^{\pm \pm}$  and  vector like lepton $F^{\pm}$ can give rise to rich phenomenological implications at the LHC. Generically, DM is searched for at the LHC in mono-$X$ searches, e.g.\ in association with one or more additional SM particles, preferably a high momentum object (jet, photon,  vector boson  etc.) radiated by the initial state quarks. 
Here, we want to highlight a few non-standard collider aspects of DM which naturally arise in our framework.\footnote{Note that charged scalars in our benchmark point are chosen to be beyond TeV, as are the vector-like fermions. This implies that they are above current sensitivities, and we can keep the discussion largely 
qualitative. }   The relevant Feynman diagram for this collider signal of DM are shown in Fig.~\ref{fyncoll}. The charged vector like fermion $F^{\pm}$, which is responsible for lepton anomalous magnetic moments, will be pair-produced at the LHC via $s$-channel $Z/\gamma$ exchange and it will further decay back to DM and SM charged leptons. This will lead to DM production in association with two charged leptons ($pp \to \ell^+ \ell^- + {E\!\!\!\!/}_{T}$)  at the LHC. This process is somehow similar to the standard slepton searches \cite{Sirunyan:2018vig,Aad:2014yka,Sirunyan:2018nwe}.  If kinematically allowed, the DM can also be produced in association with same-sign dileptons from the decay of doubly charged scalar $S^{\pm\pm}$ as shown in right panel of the Fig.~\ref{fyncoll}. The dominant production mechanism of the doubly charged scalar $S^{\pm\pm}$ at the LHC is the standard Drell-Yan process via 
$s$-channel $Z/\gamma$ exchange. It will further dominantly decay to $S^{\pm \pm} \to F^{\pm} \ell^{\pm}$ and the vector-like leptons $F^{\pm} $ decay  dominantly to DM and SM charged leptons. This will lead to  DM production in association with four charged leptons ($pp \to 2\ell^+ 2 \ell^- + {E\!\!\!\!/}_{T}$)  at the LHC. Recently, prospects of this type of DM signal with  multi-lepton signature were analyzed   in detail \cite{Chakraborti:2020zxt}. On the other hand, if  the ${\cal Z}_2$ odd charged scalars $S^{\pm \pm}$ and fermions $F^{\pm}$ are not kinematically allowed to decay to DM promptly, they will be long-lived. In this case the track originating  from long-lived charged particles can disappear at a point inside the detector. There are dedicated searches for these stable charged particles at the LHC \cite{Khachatryan:2016sfv} using signatures of long time-of-flight measurements and anomalously high energy deposits in the silicon tracker. Non-observation of any signal impose severe constraints on these  stable charged particles. Using the 13 TeV LHC data \cite{Khachatryan:2016sfv}, we find that the mass of a  (long-lived) charged vector-like fermion $F^{\pm}$  is constrained up to 550 GeV, whereas the mass limit on (long-lived) doubly charged scalar $S^{\pm \pm}$ is 660 GeV. Recently, displaced vertex and disappearing track signature for long-lived singly-charged lepton were  analyzed, see  \cite{Jana:2019tdm,Acharya:2020uwc}. Also, the prospect of discovery of long-lived doubly charged scalars was analyzed   \cite{Dev:2018kpa}. 
Thus, this model predicts several unique signals like displaced vertex signature, disappearing tracks at the collider and non-standard DM signals with multi-lepton signature.  All these signals have unique discovery prospects which can be tested in the upcoming run of the LHC or other colliders. The investigation of these collider signals is  beyond the scope of this article and shall be presented in a future work.

\section{Conclusions}\label{sec:C}
In this work, we have proposed a class of models that intercorrelates and offers a simultaneous explanation of neutrino mass,  dark matter, the long-standing puzzle of the muon anomalous magnetic moment, and the recently observed tension in the electron anomalous magnetic moment. In each of these models, the Standard Model is extended with a vector-like fermion and a set of scalars, which are odd under an added $\mathcal{Z}_2$ symmetry. A common set of these BSM states run through the loops and generates neutrino mass as well as lepton AMMs at one-loop order. If the vector-like fermions are around the TeV scale, they provide large chirality enhanced contributions required to resolve the lepton AMMs. Different signs for the muon and electron magnetic moments are arranged easily because different sets of Yukawa couplings are involved. The lightest of the neutral members of our new multiplets, either fermionic or bosonic in nature, plays the role of the dark matter, which is stabilized by the unbroken  $\mathcal{Z}_2$ symmetry. Models belonging to this class are simple in their constructions and provide a framework to unify a number of various seemly uncorrelated issues that cannot be solved  with the Standard Model. After a generic discussion, we have focused on a particular  model, and performed a detailed analysis that includes a fit to neutrino oscillation parameters as well as electron and muon AMMs, followed by a discussion of DM and collider phenomenology.

\section*{Acknowledgments}
The work of VPK was in part supported by US Department of Energy Grant Number DE-SC 0016013. WR is supported by the DFG with grant RO 2516/7-1 in the Heisenberg program.

\bibliographystyle{utphys}
\bibliography{reference}
\end{document}